\documentstyle[12pt,subfigure,graphicx,color]{article}
\begin{document}

{\large {\bf Rapid Disappearance of Penumbra-Like Features} } \\
{\large {\bf Near a Flaring Polarity Inversion Line: The Hinode Observations} } \\

\vspace {0.5cm}
\centerline{B. Ravindra$^{1}$ and Sanjay Gosain$^{2,3}$}

\centerline{$^{1}$Indian Institute of Astrophysics, Koramangala, Bangalore - 560034, India}
\centerline{$^{2}$Udaipur Solar Observatory, Dewali, Badi Road, Udaipur - 313001, India}
\centerline{$^{3}$National Solar Observatory, 950 N Cherry Avenue, Tucson 85719 AZ, USA}

\vspace{1.0cm}
\noindent{Abstract:}
We present the observations of penumbra like features (PLFs) near a
polarity inversion line (PIL) of
flaring region. The PIL is located at the moat boundary of active
region (NOAA 10960). The PLFs appear similar to sunspot penumbrae
in morphology but occupy small area, about 6$\times$10$^{7}$~km$^{2}$,
and are not associated with sunspot or pore. We observed a rapid
disappearance of the PLFs after a C1.7 class flare, which occurred
close to the PIL. The local correlation tracking (LCT) of these
features shows presence of horizontal flows directed away from the
end-points of the PLFs, similar to the radial outward flow found
around regular sunspots, which is also known as the moat flow. Hard X-ray
emission, coincident with the location of the PLFs, is found in
RHESSI observations, suggesting a spatial correlation between the
occurrence of the flare and decay of the PLFs. Vector magnetic field
derived from the observations obtained by Hinode spectro-polarimeter
SOT/SP instrument, before and after the flare, shows a significant
change in the horizontal as well as the vertical component of the
field, after the flare. The weakening of both the components of
the magnetic field in the flare interval suggests that rapid
cancellation and/or submergence of the magnetic field in PLFs
occurred during the flare interval.

\section{Introduction}

\noindent
Well developed sunspots have umbra surrounded by  elongated filamentary
type of structures called penumbral filaments. The magnetic field in
the umbra is mostly vertical to the solar surface while in the penumbra
it is horizontally inclined. Further, the inclination of the penumbral
magnetic field shows an azimuthal variation, with channels of more
horizontal field lying in between the channels of more vertical fields.
This alternating inclination of penumbral field is also called as
``uncombed penumbra" and is well established from the past observations
(Solanki and Montavan, 1993). The more horizontal component of the
uncombed field is observed to harbor most of the Evershed flow.
Penumbrae form around pores  abruptly with an increase in the magnetic
field inclination at the boundary(Tildesley and Weiss, 2004). Penumbral
filament  formation and Evershed flow occurs nearly simultaneously
in pores (Leka and Skumanich, 1998), i.e., as soon as the penumbrae
form the Evershed flow can be observed. It is found that a magnetic
flux threshold above which penumbra forms is about
1-1.5$\times$10$^{20}$~Mx (Leka and Skumanich, 1998).

Once the penumbra is formed it survives from hours to days exhibiting
a variety of small scale structures like dark cores inside penumbral
filaments, apparently propagating twisting structures, moving magnetic
features to name a few.  Evershed flow (Evershed, 1909),a mass flow
directed radially outwards from the sunspot along the penumbra is
observed in the photosphere,while inverse Evershed flow, directed
towards sunspot is observed in the chromosphere.The actual lifetime
of the penumbra depends upon the life of the sunspot, when the
spot decays it gradually loses its penumbra, becomes a
``naked" sunspot and finally fragments into pores. Moving
Magnetic Features (MMFs: Harvey and Harvey, 1973)
are proposed as one of the candidates for the decay of
sunspots. Once the sunspot decays, naturally the penumbra
disappears. Decay of sunspots is studied in detail by
Martinez Pillet (2002) and found that not all MMFs are related
to the decay of sunspots. Bellot Rubio et al. (2008) have found
that the absence of Evershed flow in penumbral field lines can
raise the penumbral filaments to the chromosphere that can cause
the disappearance of the penumbra at the photospheric level.Large
flares can also cause the disappearance of the penumbra at the
outer boundary of the sunspots(Wang et al. 2004;
Sudol and Harvey 2005). It is believed that the field becomes
more vertical at the outer edge of sunspot as the overall active
region field collapses towards the flaring PIL(Gosain 2012),
which causes disappearance of penumbrae.

In this paper, we report on new type of penumbra like features (PLFs)
that are not associated with sunspot. Further, we use LCT method to
study the horizontal flow patterns around these structures. We then
study the evolution of PLFs in the photosphere, chromosphere and
in vector magnetograms using the space based data obtained from
Hinode/SOT and find that the structure rapidly disappears during
flare. In the forthcoming sections, we present details of the
G-band, Ca~II~H, spectro-polarimetric data and its analysis. This
is followed by the observational results on the disappearance of
PLFs. Possible explanation for the flare related disappearance of
PLFs is discussed in the last section.

\vspace{0.3cm}
\section{Data and Analysis}
\noindent
The space based telescope, Solar Optical Telescope
(SOT: Tsuneta et al., 2008) onboard Hinode, obtains images of
the Sun at an unprecedented resolution of about 0.2$^{\prime\prime}$ of
the photosphere and chromosphere. The broadband filter imager
(BFI) instrument on the SOT produces images in several
wavelengths with the 3-8~\AA~bandwidth filters. In this
study, we have used filtergrams observed at 4305~\AA~(G-band) and
3968.5~\AA~(Ca~II~H). We have used the filtergrams observed in
these wavelengths from June 06, 2007at 15:00~UT to June 07,
2007 05:00~UT. The data are corrected for dark current,pixel-to-pixel
gain variations, hot and dead pixels. The Ca~II~H data are corrected
for the wavelength dependent pixel size with the G-band as
a reference. These calibrated data sets are rigidly aligned
with the first image in the time series using a 2-dimensional
cross-correlation algorithm. The alignment of the images is
good to a sub-pixel accuracy. The aligned data sets are passed
through a subsonic filter with a cutoff value of
4 km~s$^{-1}$ to remove the effect of acoustic oscillations
(Title, et~al. 1989).

The spectro-polarimeter (SOT/SP: Ichimoto, et~al. 2008) instrument
is one of the back-end instruments of
SOT onboard Hinode satellite, makes maps of the active region
by spatial scanning and obtains Stokes I, Q, U and V spectra
in Fe I 6301.5 and 6302.5~\AA~lines. A fast-mode raster scan
with 980 steps at a step size of 0.295$^{\prime\prime}$ along
the scanning direction and 0.317$^{\prime\prime}$/pixel resolution
along the slit direction made a raster image with a field-of-view of about
290$^{\prime\prime}$$\times$162$^{\prime\prime}$.
The Stokes signals are calibrated using standard SolarSoft pipeline
for SP. The magnetic fields are obtained by using an inversion
scheme based on the Milne-Eddington algorithm (Skumanich and
Lites 1987; Lites and Skumanich 1990). The ambiguity in the
azimuth is resolved based on the minimum energy algorithm
(Metcalf, 1994; Leka et~al. 2009). Later, the magnetic field
vector has been transformed to the disk center
(Venkatakrishnan et~al., 1989). The resulting vertical
(B$_{z}$) and transverse (B$_{t}$) magnetic fields have measurement
errors of 8~G and 30~G, respectively.These magnetograms are used
to study the magnetic field parameters of the PLFs.

\section{Results}

\begin{figure*}
\begin{center}
\includegraphics[width=100mm]{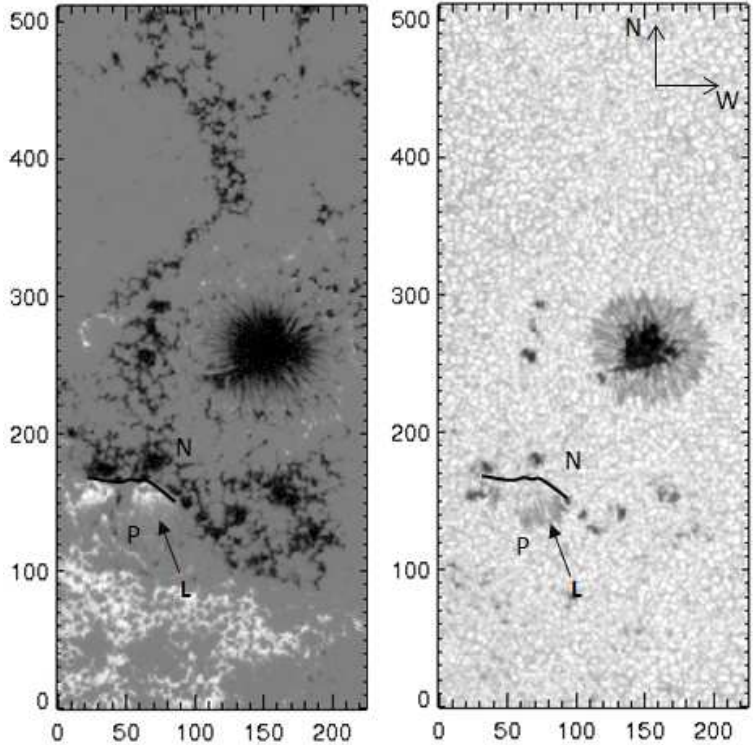} \\
\end{center}
\caption{A map of line-of-sight component of the magnetic field (left) and the 
corresponding continuum image (right) obtained from spectro-polarimeter onboard 
Hinode  on June 06, 2007 at 22:57~UT. The arrow indicates the position of PLFs. 
P and N represents the positive and negative polarity regions.
A dark line between  negative and positive polarity represents the position of 
the polarity inversion line. The scales are in unit of pixel.}
\label{fig:2}
\end{figure*}

We focus our study on the small scale penumbra like feature located
at the moat boundary of the active region NOAA 10960 observed from
6-7th June, 2007. The AR NOAA 10960 consists of two large sunspots
of negative polarity, a plage region of positive polarity, and
several pores of either polarity(http://www.solarmonitor.org).
Two sunspots were aligned along the East-West direction at a latitude
of 6$^\circ$ in the southern hemisphere. During
our observations of event, the active region was close to the disk
center (S06E05). Active Region NOAA 10960 produced  several
B and C class flares with a very few M class flares and the
largest one was observed on June 04, 2007 which was of M8.9 class.
A few of these events were studied in detail  by
Srivastava et~al. (2010) and Kumar et~al. (2010), where
they have observed the sunspot rotation and kink instability
that lead to the initiation of the flare. During Hinode observations,
there was one C1.7 class flare on June 06, 2007 at23:31~UT and it
occurred close to the polarity inversion line (PIL marked in
Figure \ref{fig:2}). On June 07, 2007  between 00:00 to 05:00~UT
there were two B-class flares and their magnitude was B7.6 and B6.6
peaking at 00:45 and 01:40~UT respectively. This event was well
observed by the Hinode/SOT at high cadence (21~sec.) in Ca~II~H
band with some small data gaps in-between.It was also observed in
G-band at a slower cadence of about 100 sec.

\begin{figure*}
\begin{center}
\includegraphics[width=100mm]{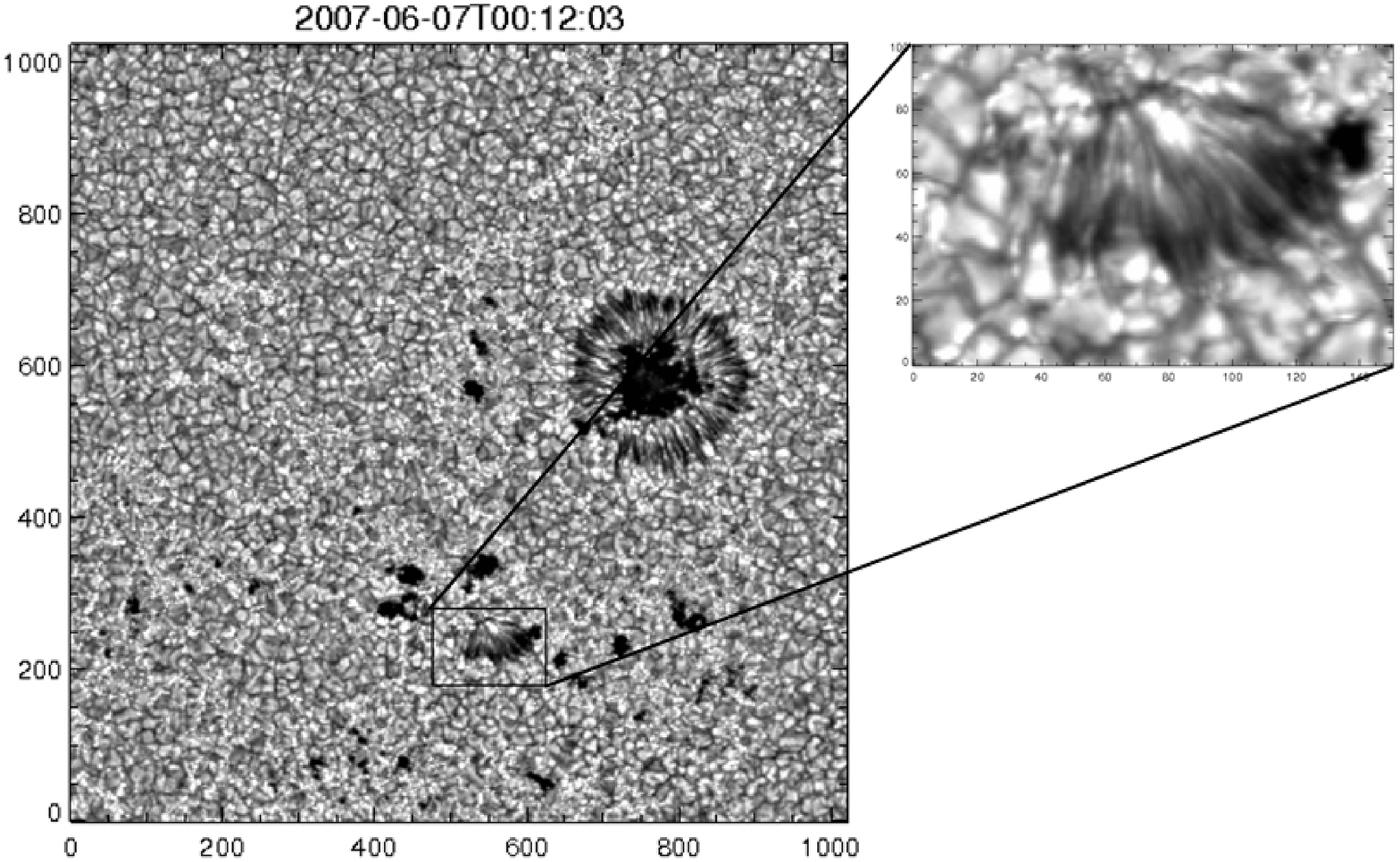} \\
\includegraphics[width=100mm]{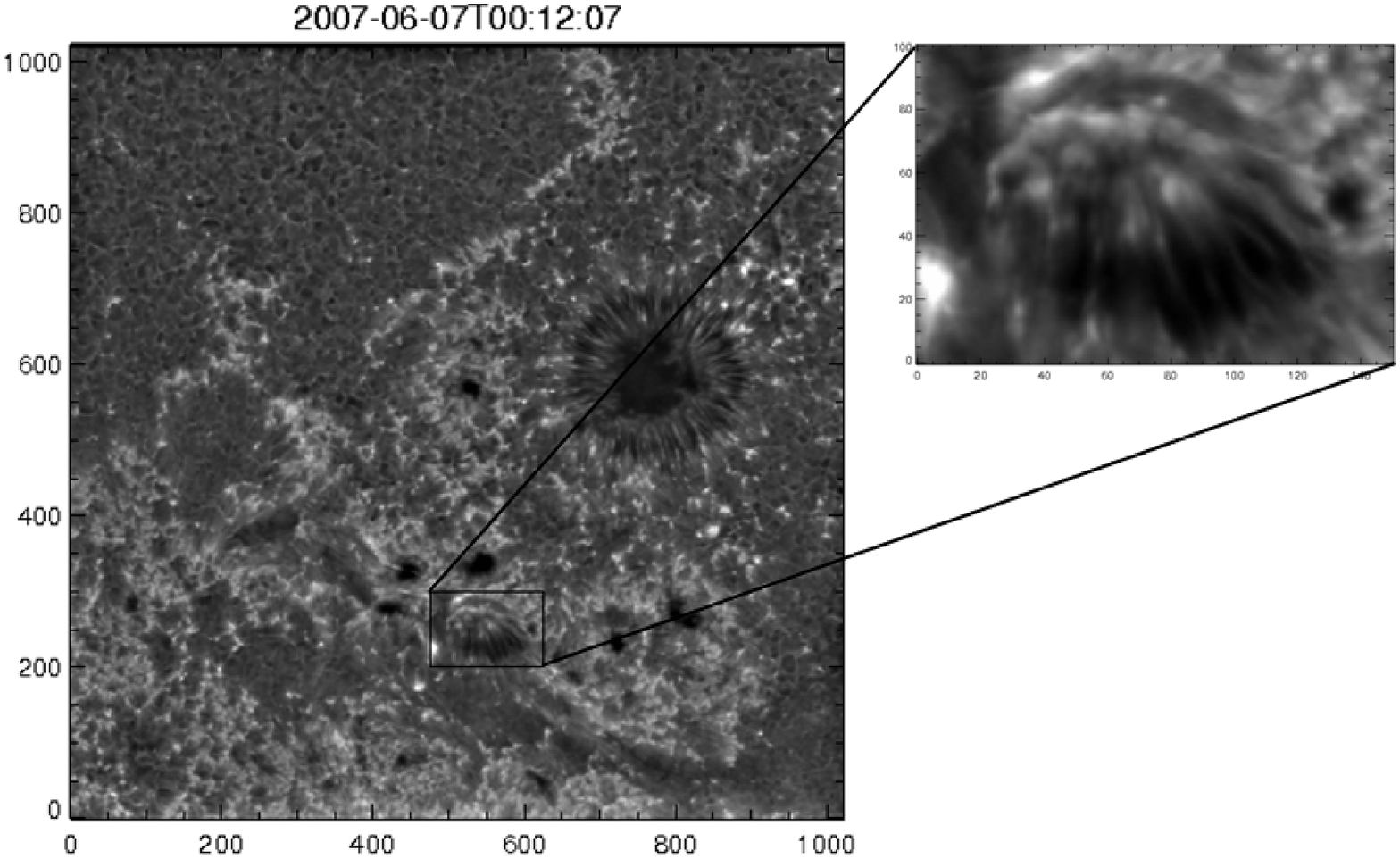}\\
\end{center}
\caption{Top: A sample image is obtained from Hinode/SOT using G-band
filter. The boxed region shows the region of interest. A magnified view
of the region of interest is shown on the right side of it. Bottom:
Same as top side image but obtained in the wavelength of
3968~\AA~of Ca~II~H. The scales are in unit of pixel.}
\label{fig:1}
\end{figure*}

Figure \ref{fig:1} (top) and (bottom) shows the sunspot in the active region
at the photospheric and chromospheric level. Penumbra like features (PLFs) are
observed close to the sunspot and it is marked with a box. An enlarged portion 
of boxed region is shown next to it. Close to the PLFs region, three pores are 
visible which are quite separated from the PLFs and are not like the pores which 
form rudimentary penumbra. The same structure is also visible in the chromospheric 
image taken in Ca~II~H wavelength. The region of interest is shown in box is 
magnified and displayed along side. Clearly, in the magnified image one can 
see detailed view of the PLFs. They exhibit about 10$^{\prime\prime}$ linear 
size and appear very much like the sunspot penumbra. The AR 10960 was a 
fast evolving region and started to disrupt from June 04, 2007
(see: http://solar-b.nao.ac.jp/QLmovies/movie$\_$sirius/2007/06/04/hsc$\_$ql20070604$\_$e.shtml). 
The formation process of the PLFs could not be clearly identified as the cadence of the 
observations is intermittent and the field-of-view of Hinode SOT is also limited.
However, here we concentrate more on two aspects (i) brief investigation of the properties 
of the new PLFs identified in Hinode G-band observations, and (ii) their evolution in 
relation to recurrent flares. The search for the origin or formation of PLFs would 
require a fully dedicated high resolution observational campaign, which we plan to do 
in near future either with Hinode SOT or ground based high resolution telescopes 
equipped with adaptive-optics. 

Interestingly,
the PLFs are located next to the polarity inversion line (PIL) with PLFs filaments 
aligned nearly parallel to the PIL (Figure \ref{fig:2}).
In Figure \ref{fig:2} the left side image is the line-of-sight magnetogram and the
right side image is the continuum image. Both the images are obtained from SOT/SP.
A thin dark line in-between the positive and
negative polarity is shown to represent the position of the PIL.
Close to this PIL numerous B and C-class flares were observed over a
couple of days. The PLFs (location `L', shown by an arrow in Figure \ref{fig:2})
has positive polarity and is located close to the PIL.
Here, the magnetic field in the lower part of PLFs (close to the arrow-head) is
more inclined leading to smaller signal in line-of-sight component of the magnetic 
fields. In the magnetogram, it is very clear that the region PLFs is located in the 
moat boundary of the negative polarity sunspot.

\subsection{Evolution of the PLFs}

\begin{figure*}
\begin{center}
\includegraphics[width=40mm]{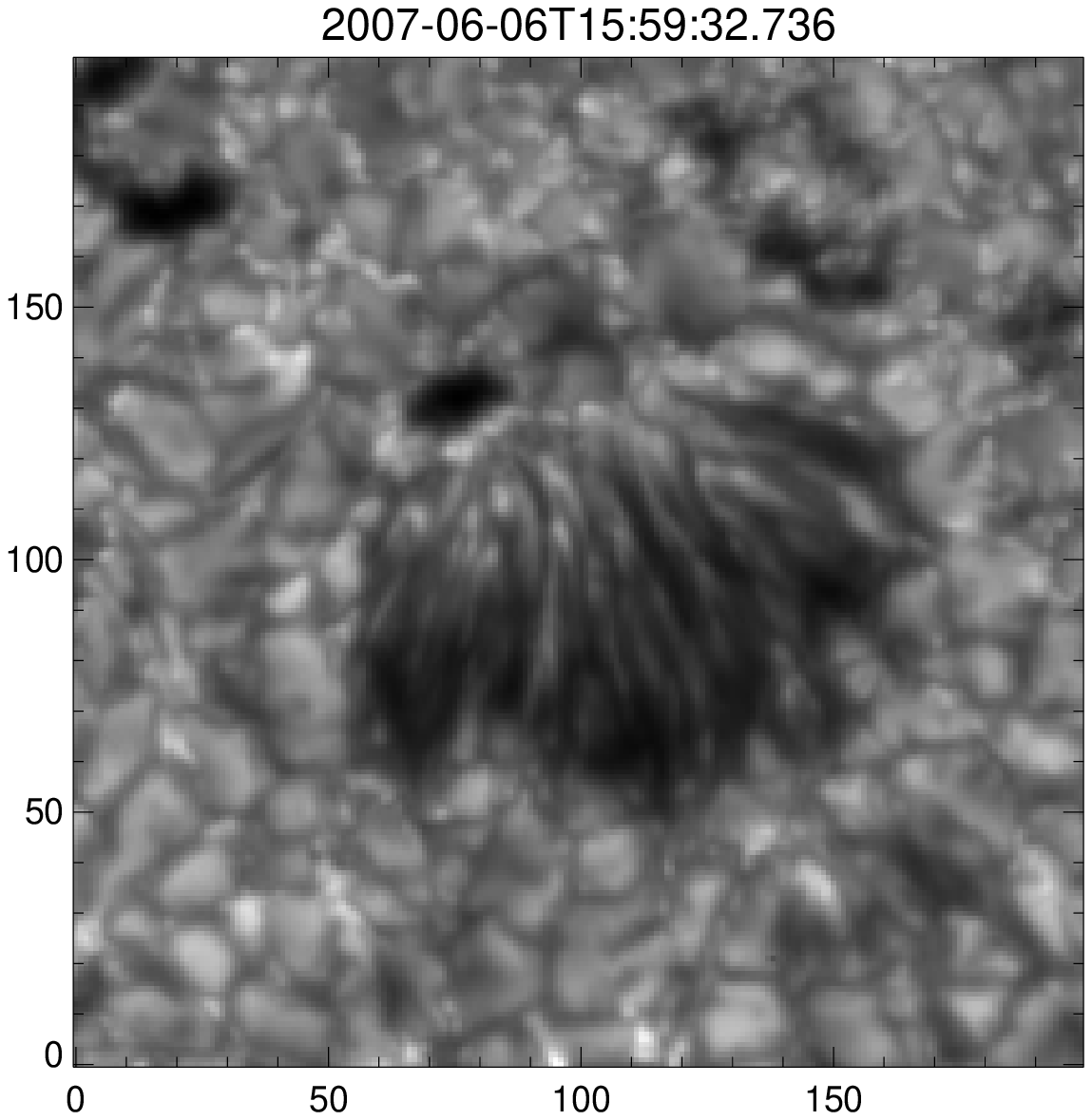}\includegraphics[width=40mm]{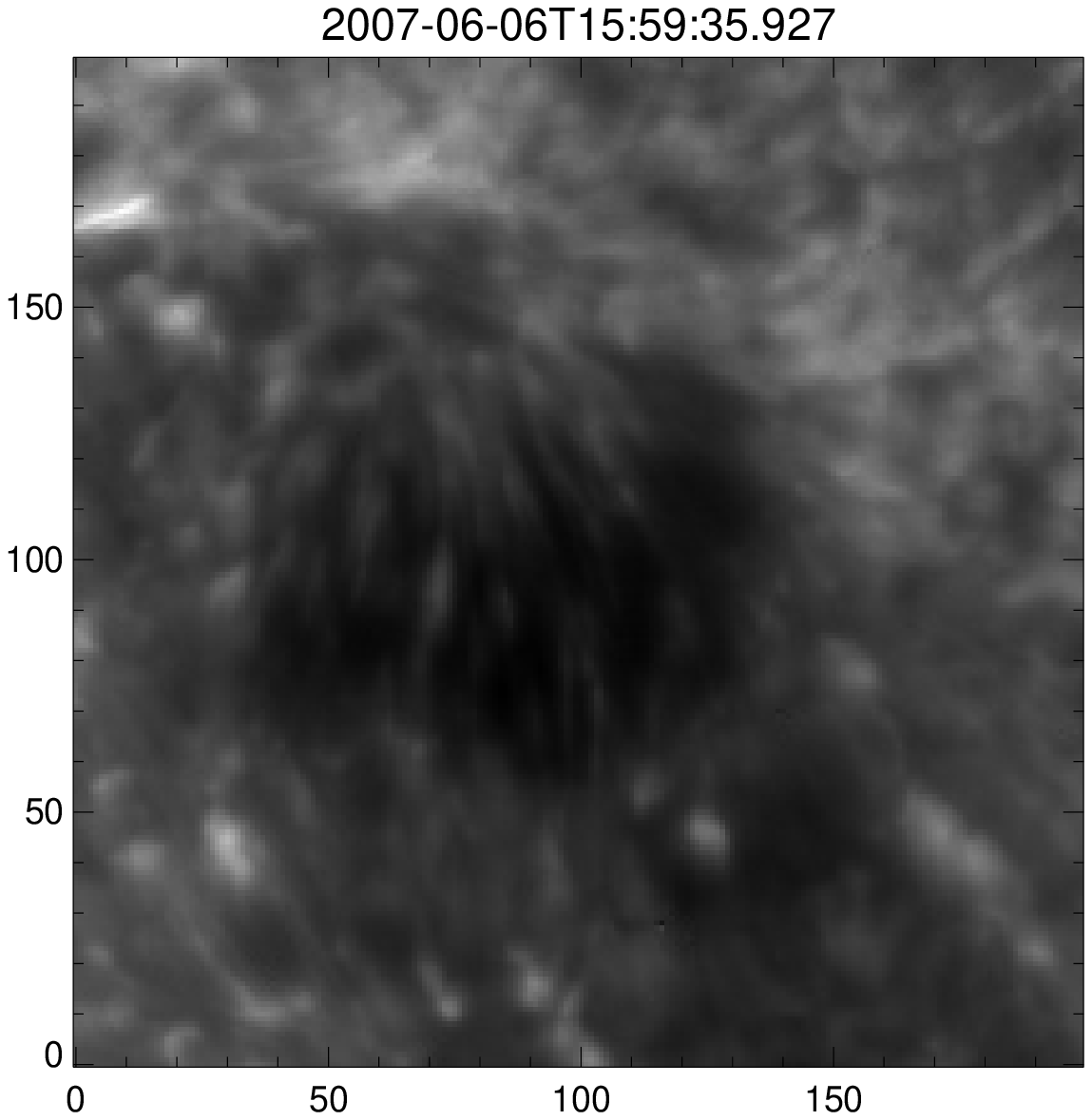}\includegraphics[width=40mm]{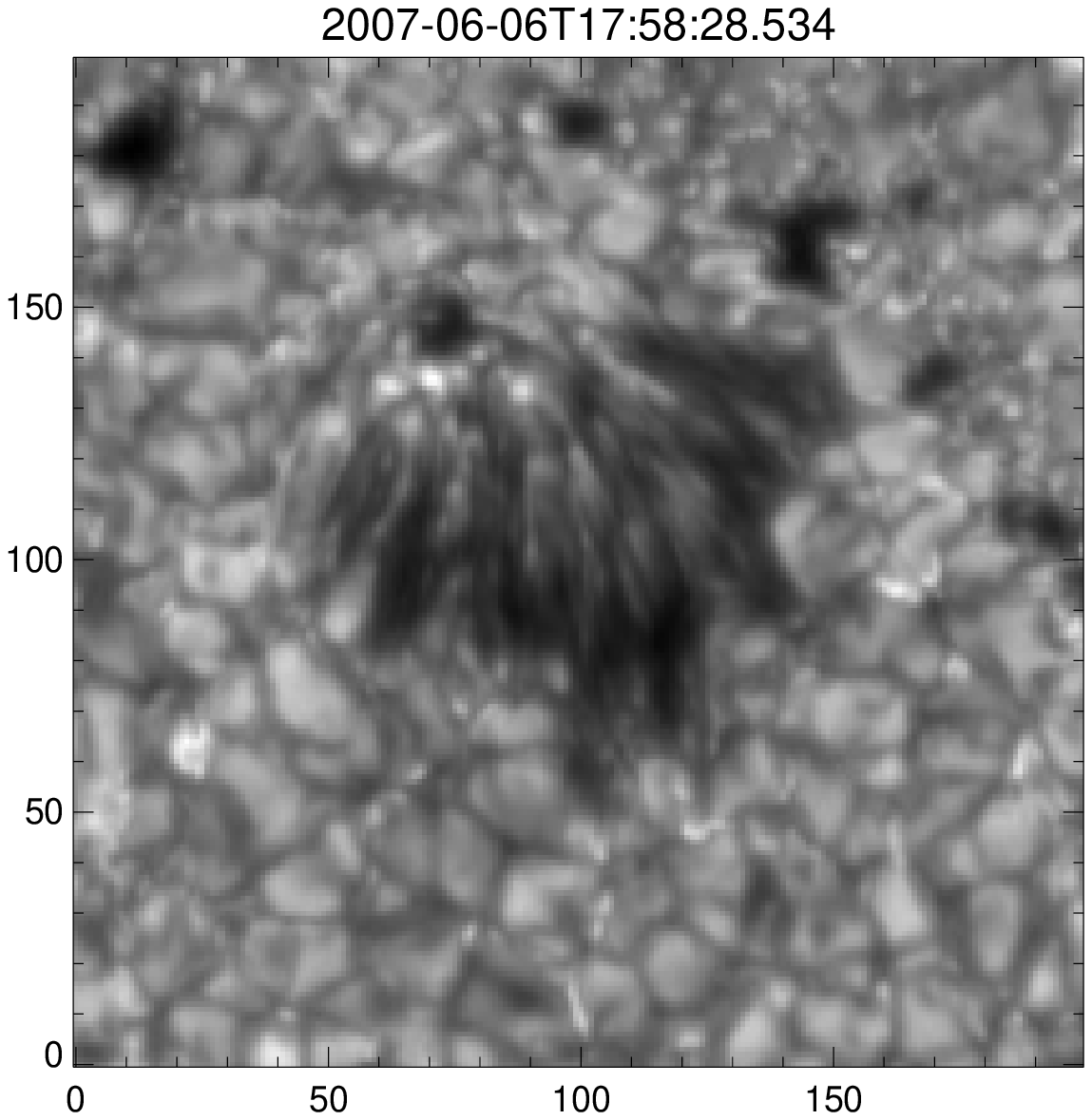}\includegraphics[width=40mm]{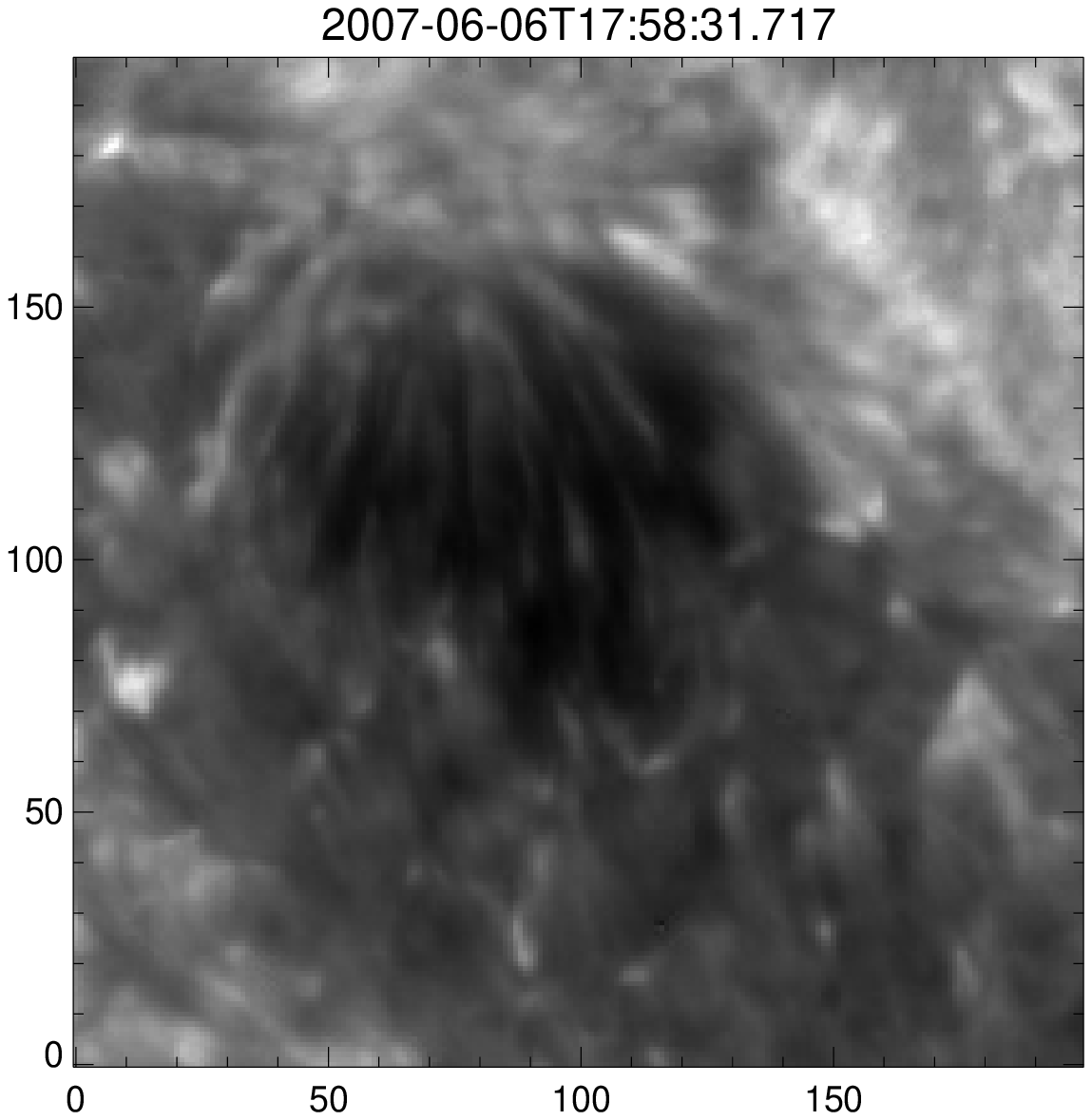} \\
\includegraphics[width=40mm]{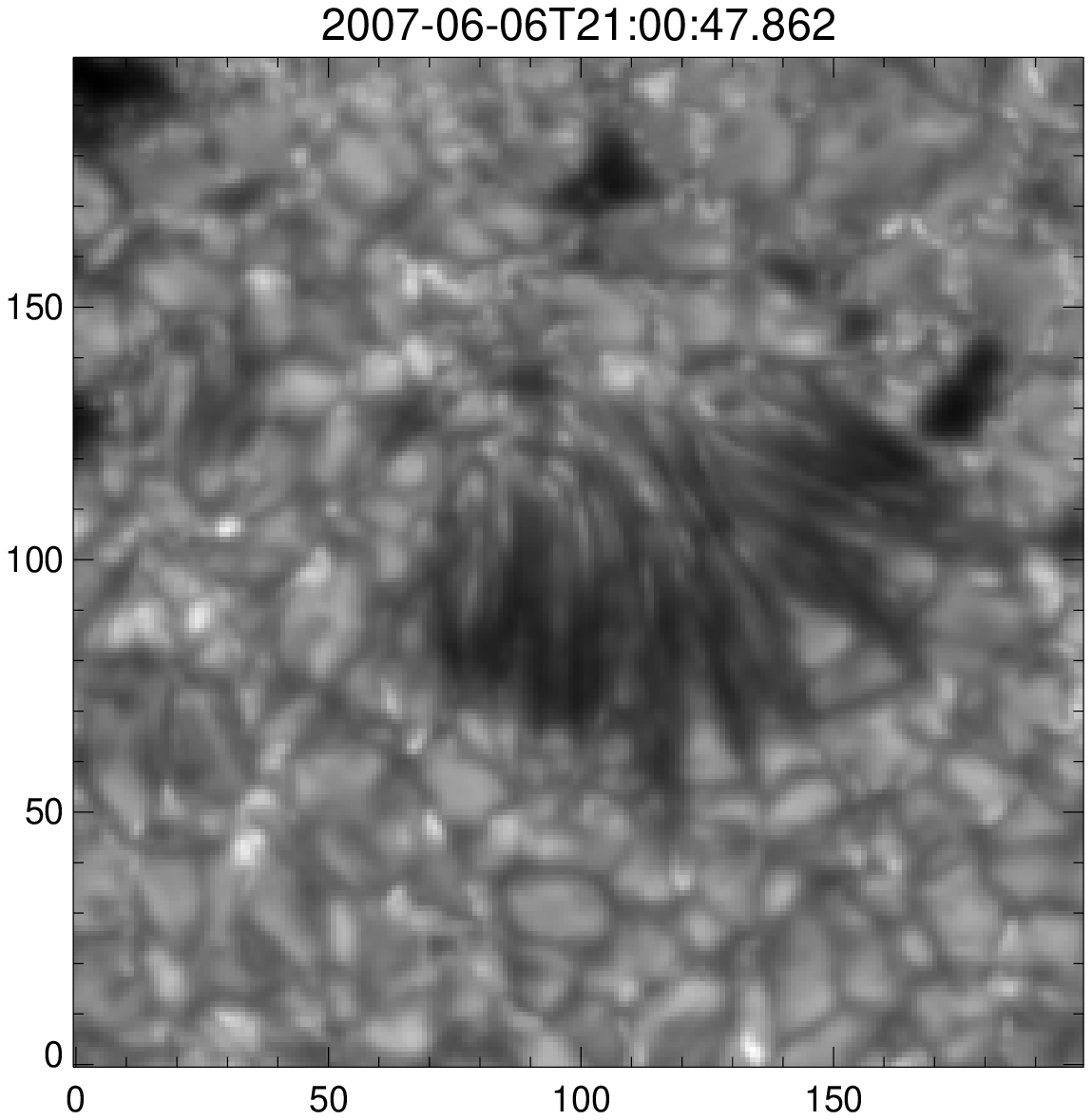}\includegraphics[width=40mm]{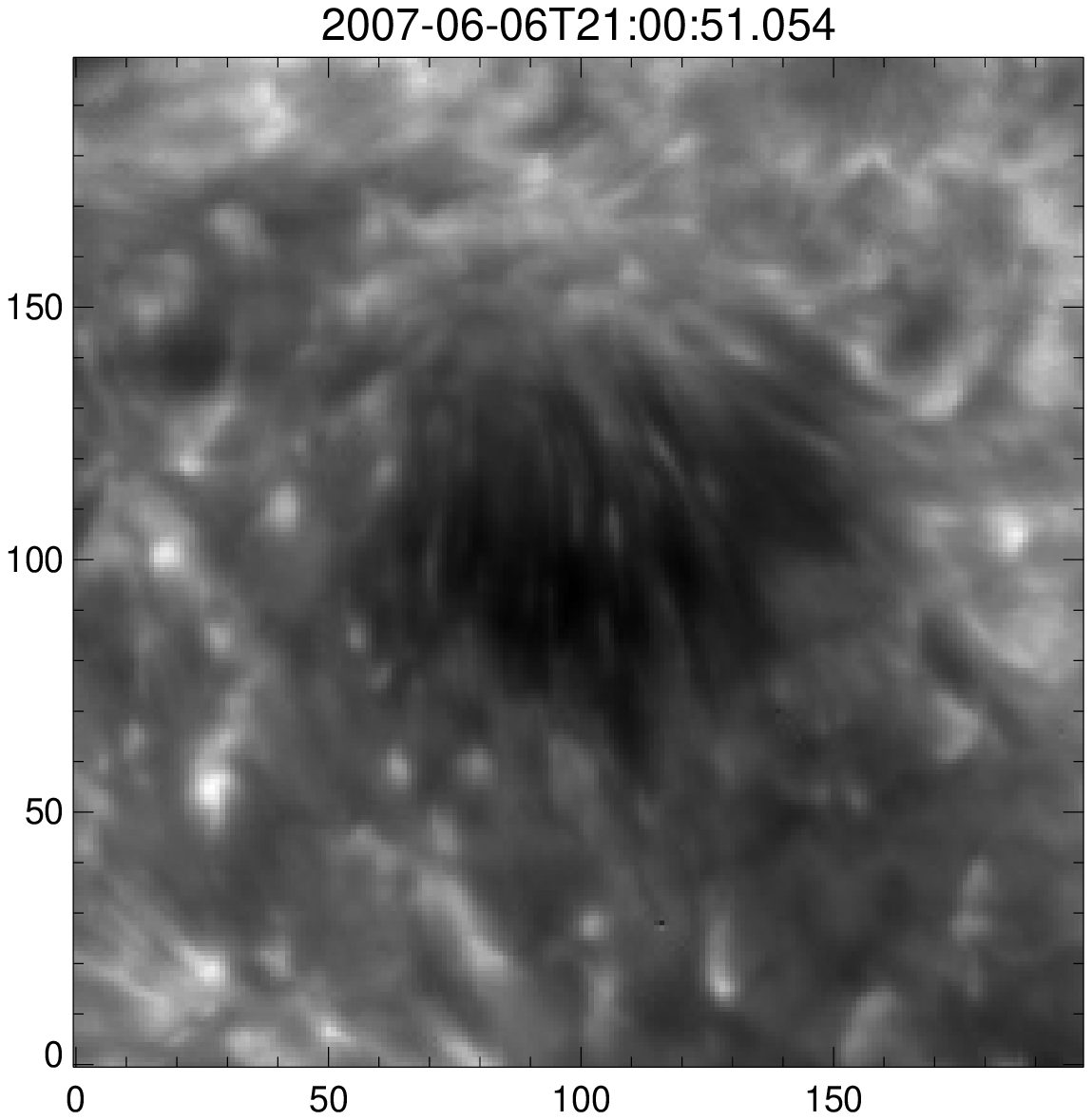}\includegraphics[width=40mm]{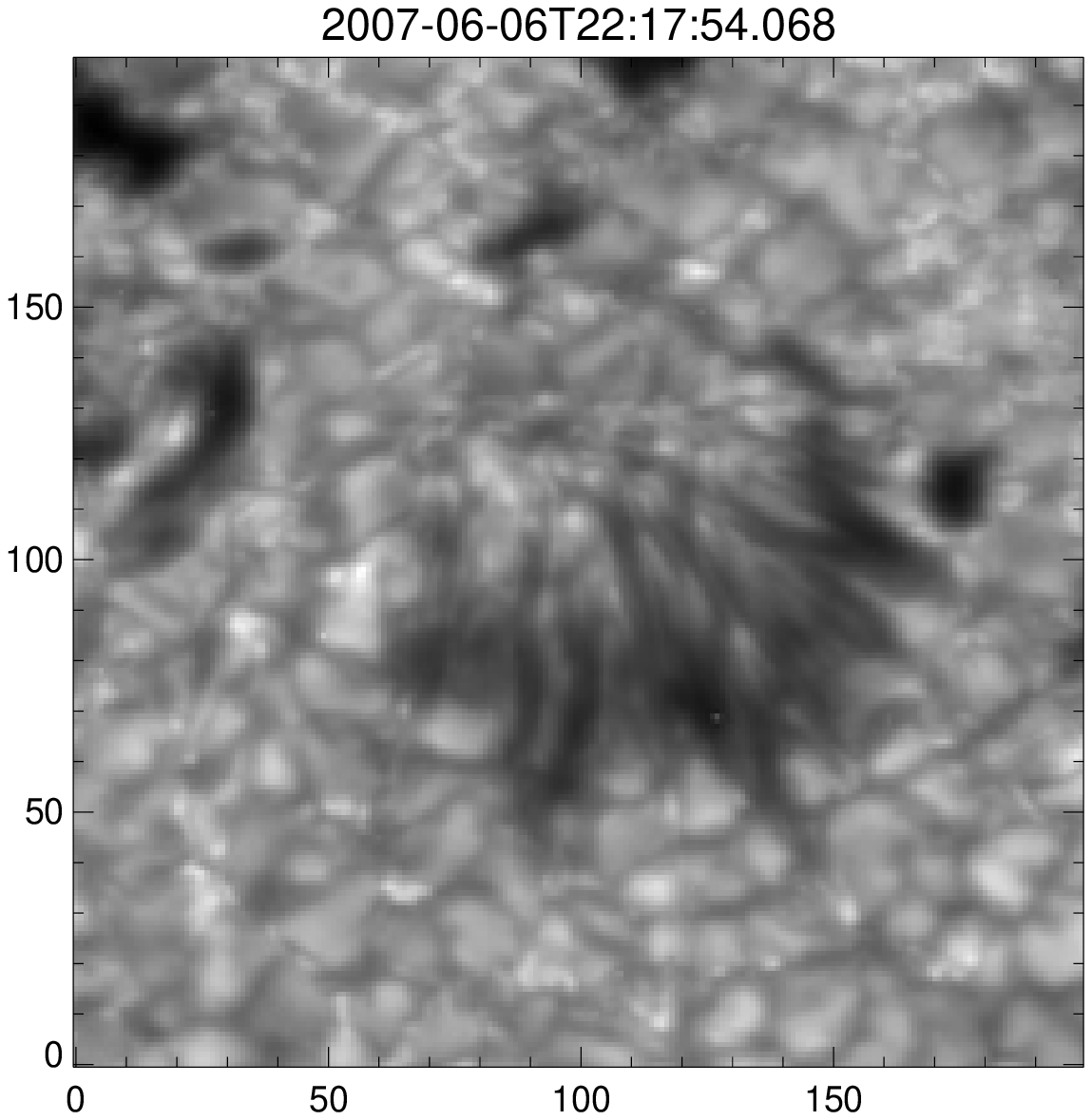}\includegraphics[width=40mm]{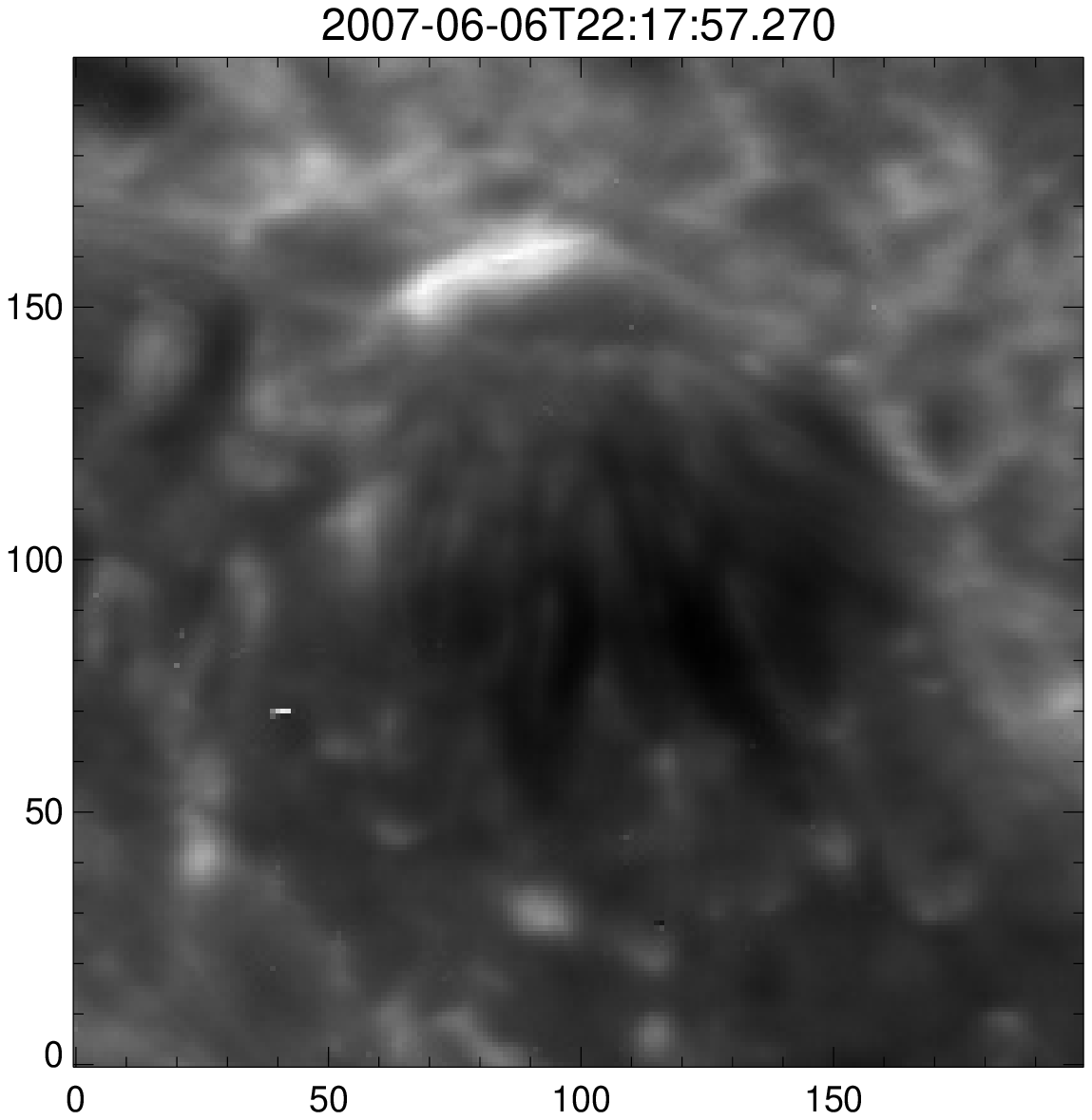} \\
\includegraphics[width=40mm]{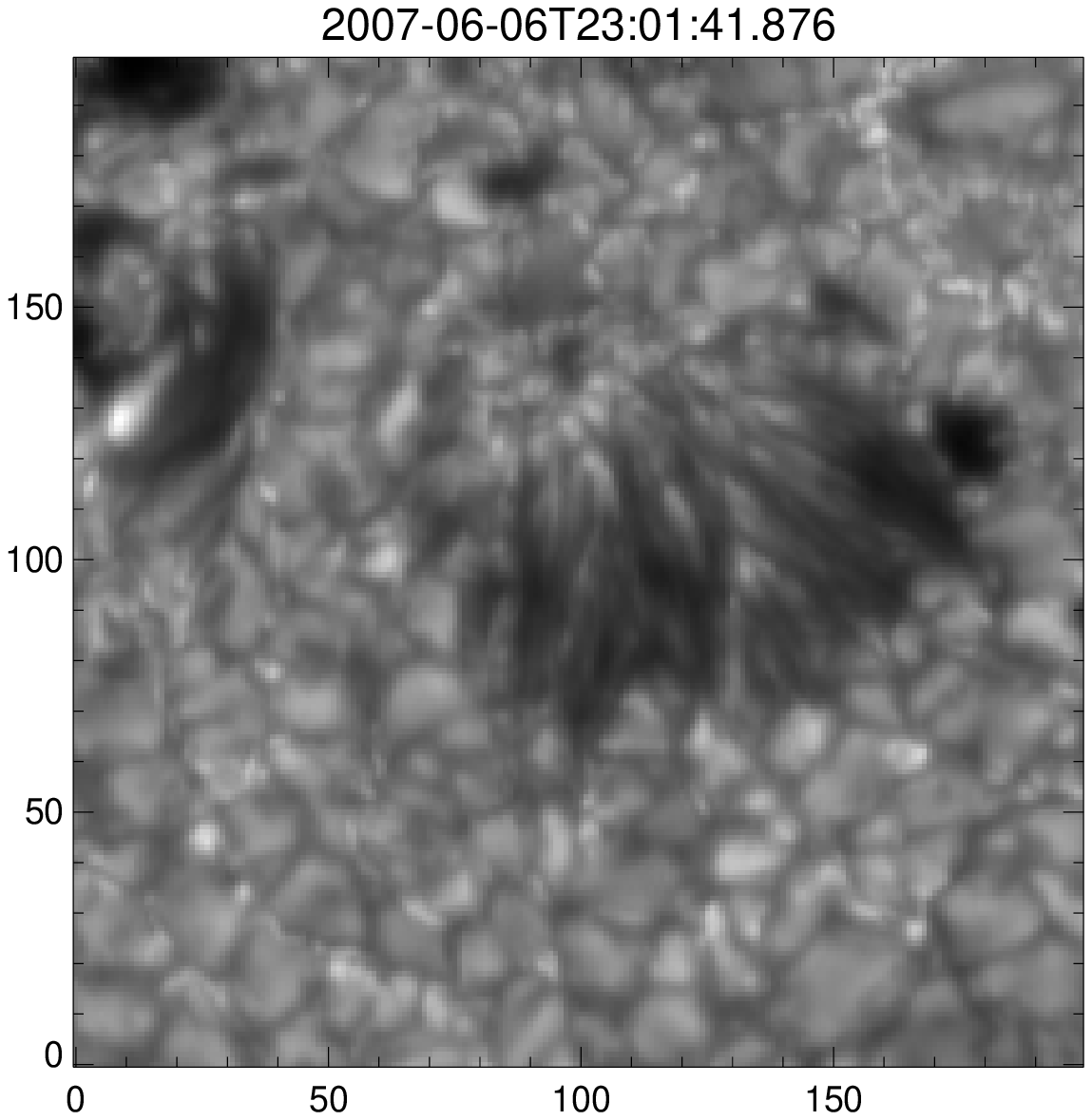}\includegraphics[width=40mm]{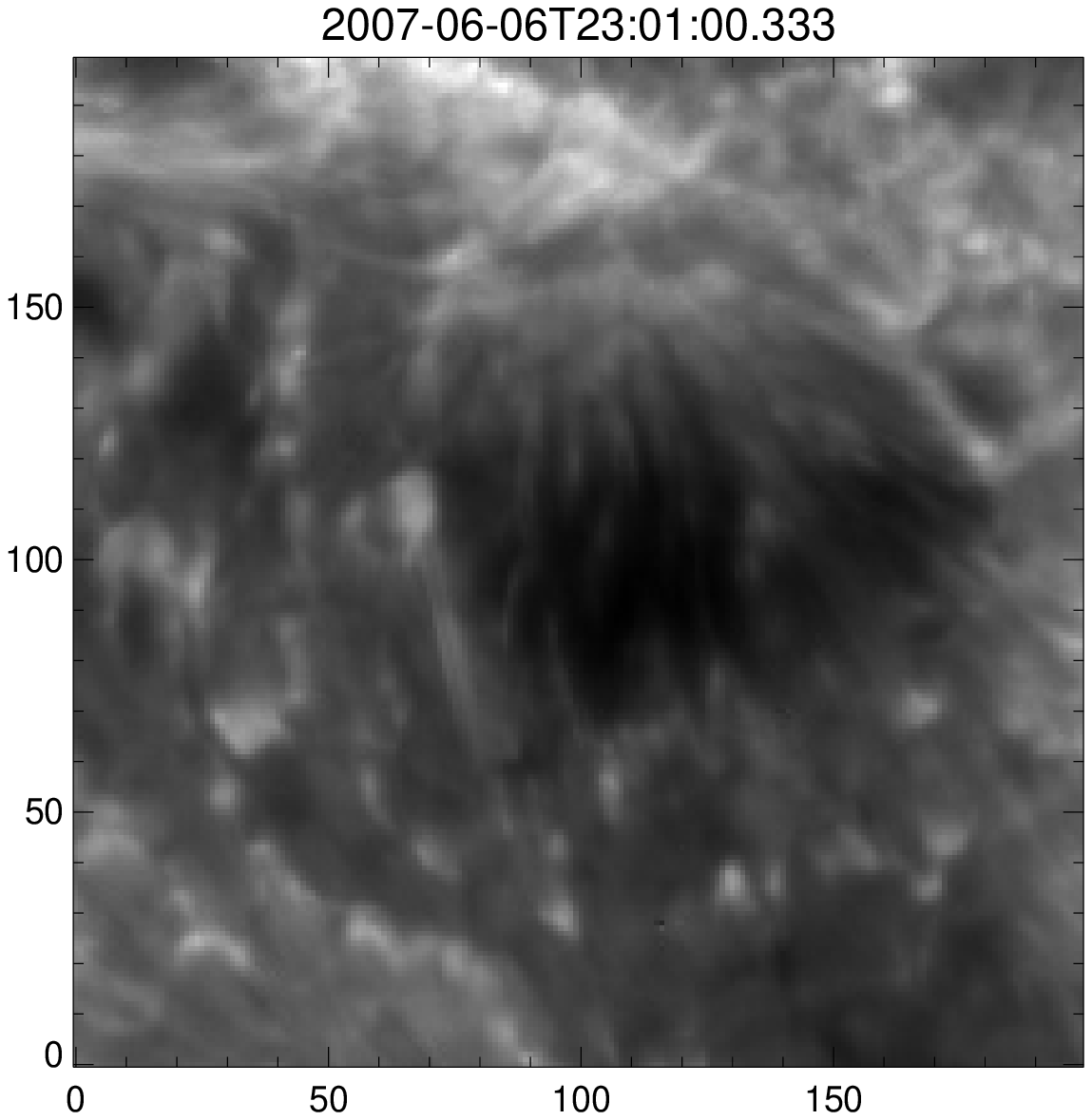}\includegraphics[width=40mm]{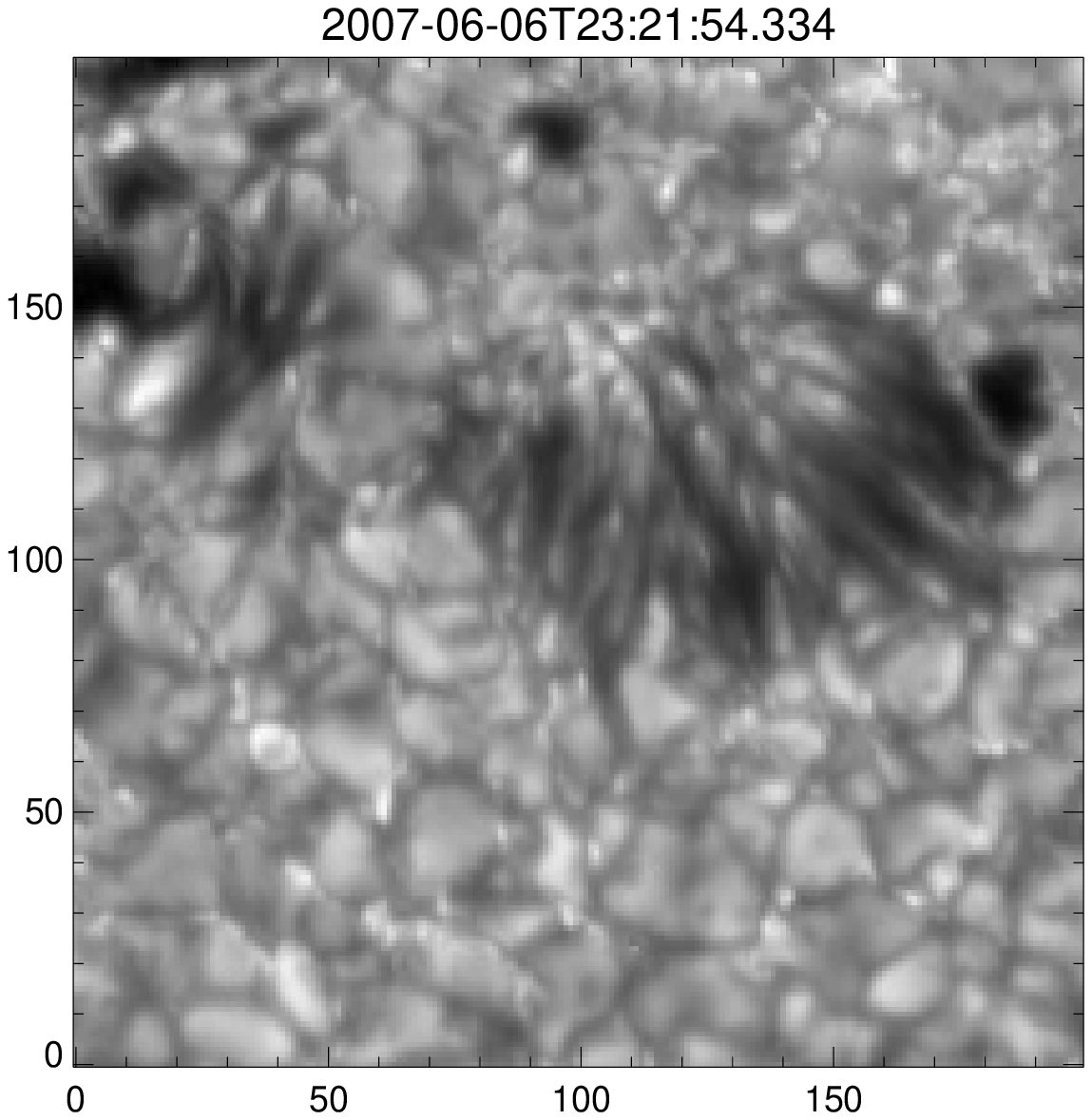}\includegraphics[width=40mm]{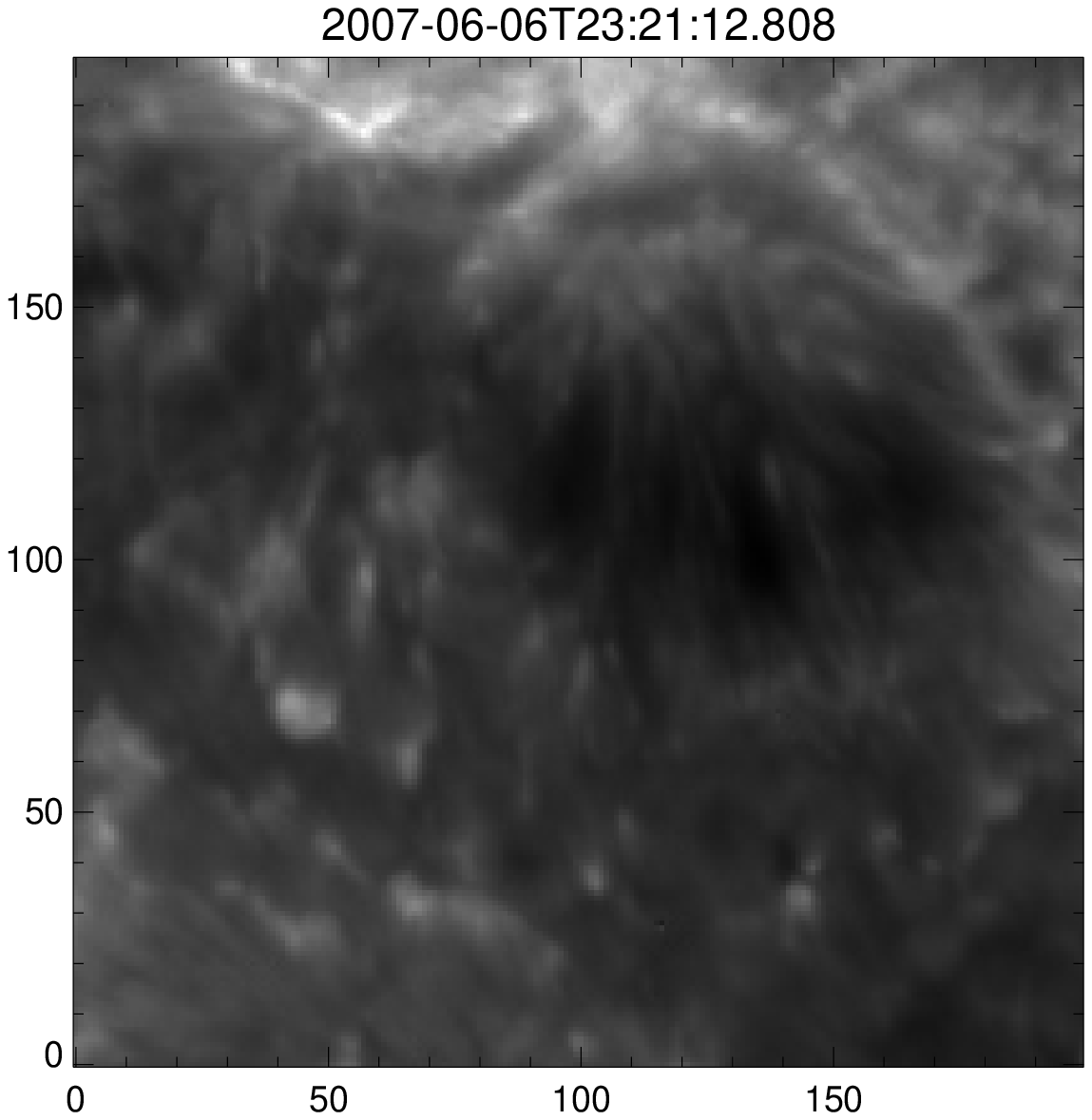} \\
\includegraphics[width=40mm]{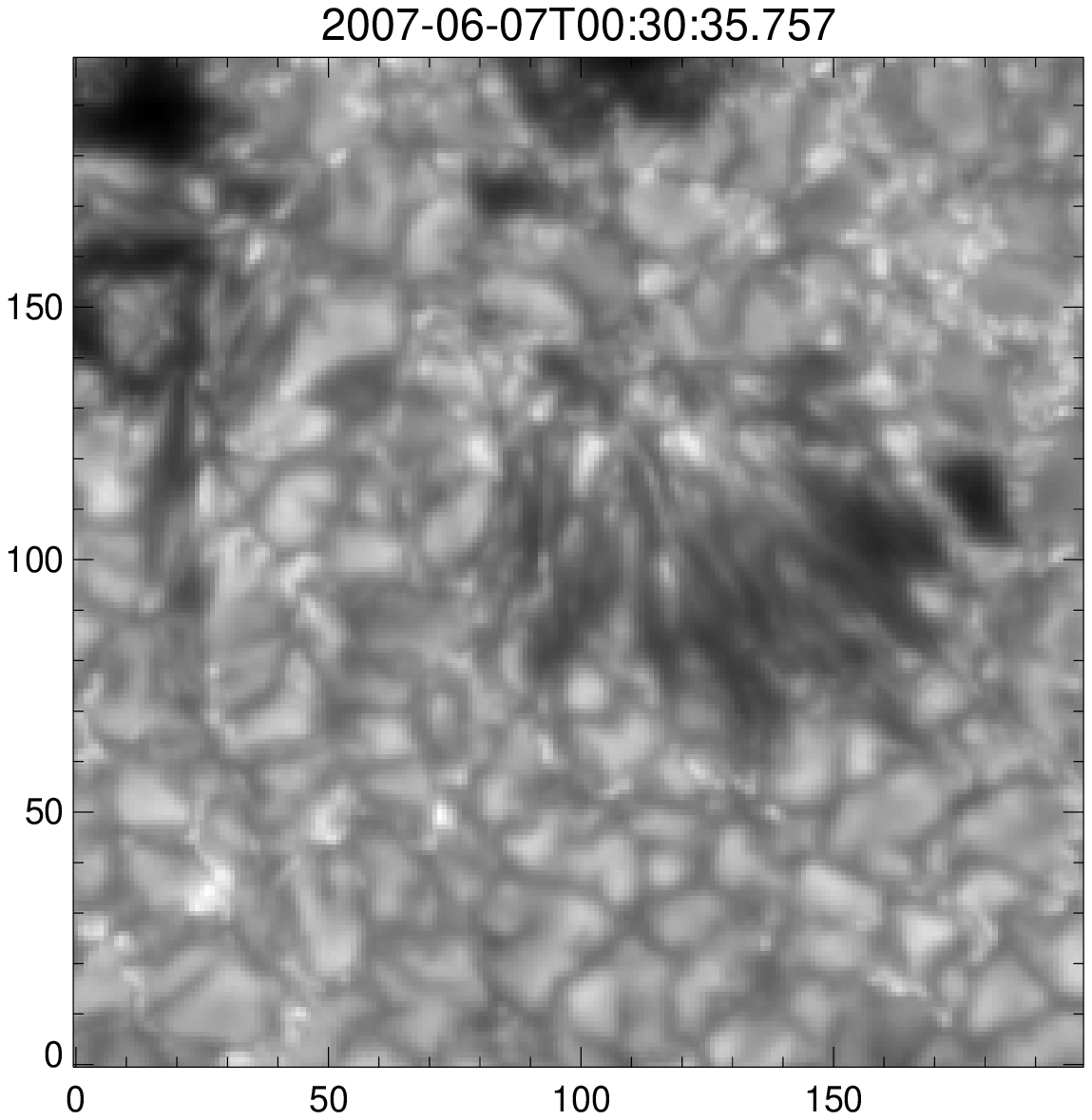}\includegraphics[width=40mm]{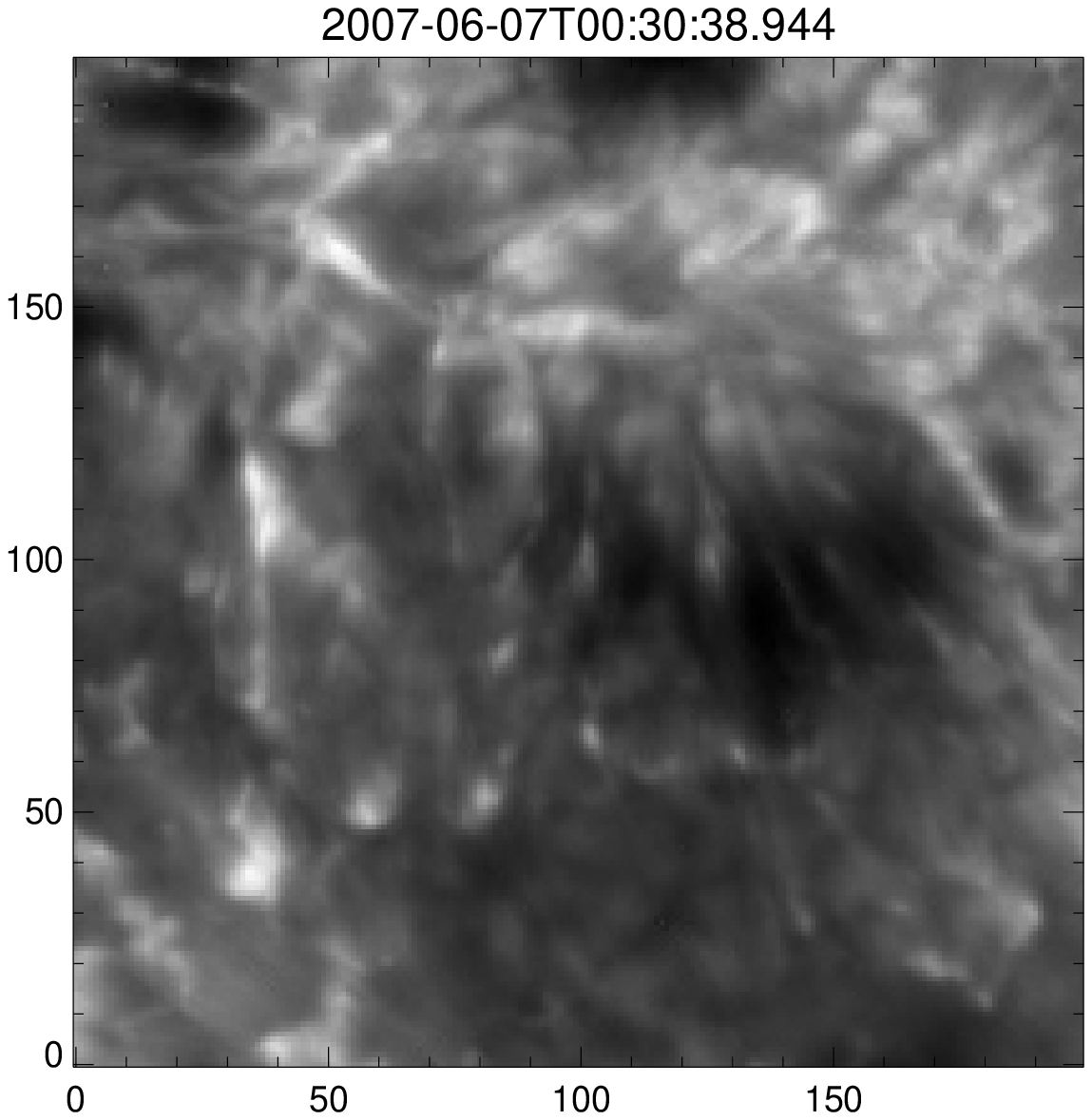}\includegraphics[width=40mm]{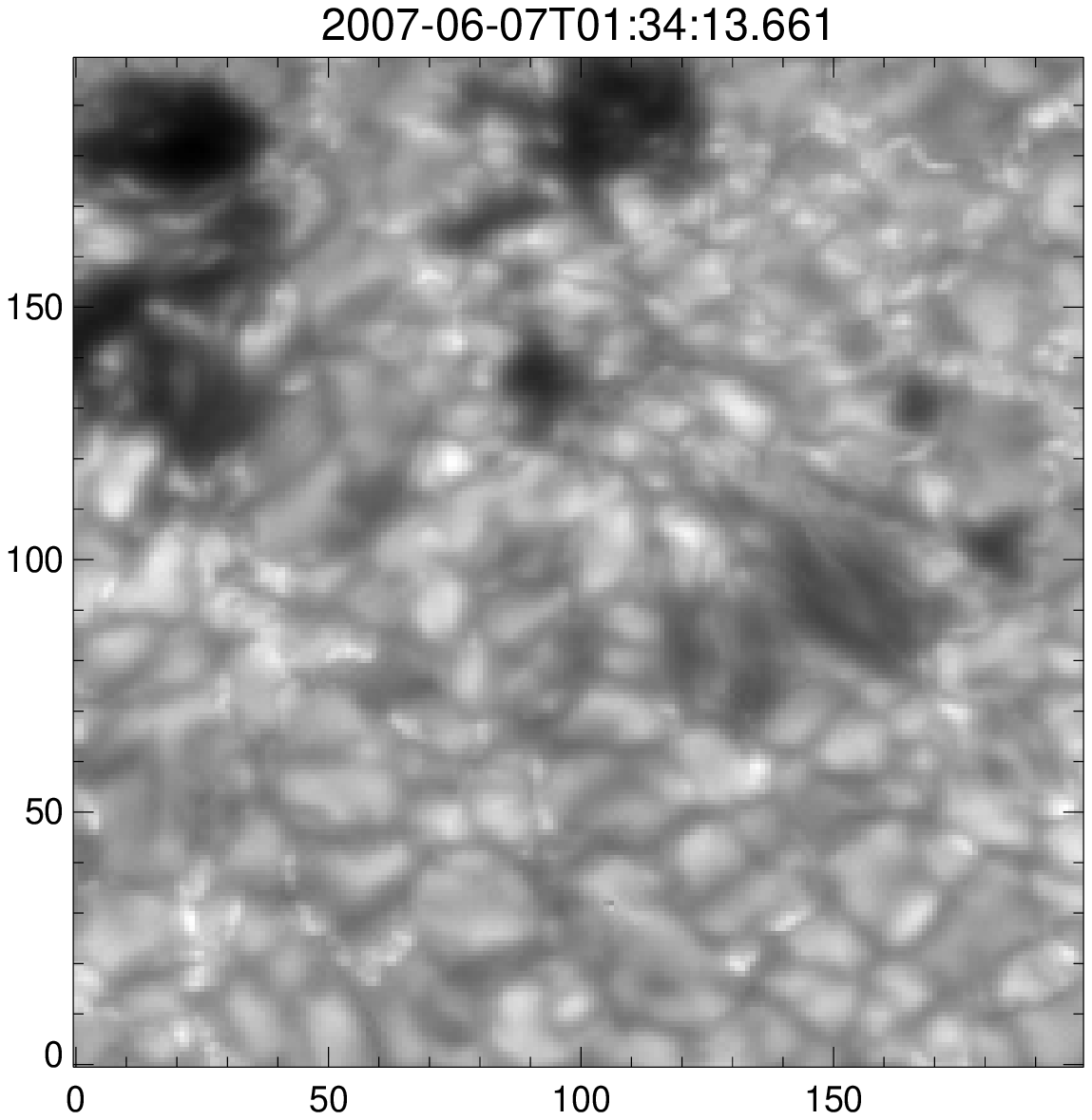}\includegraphics[width=40mm]{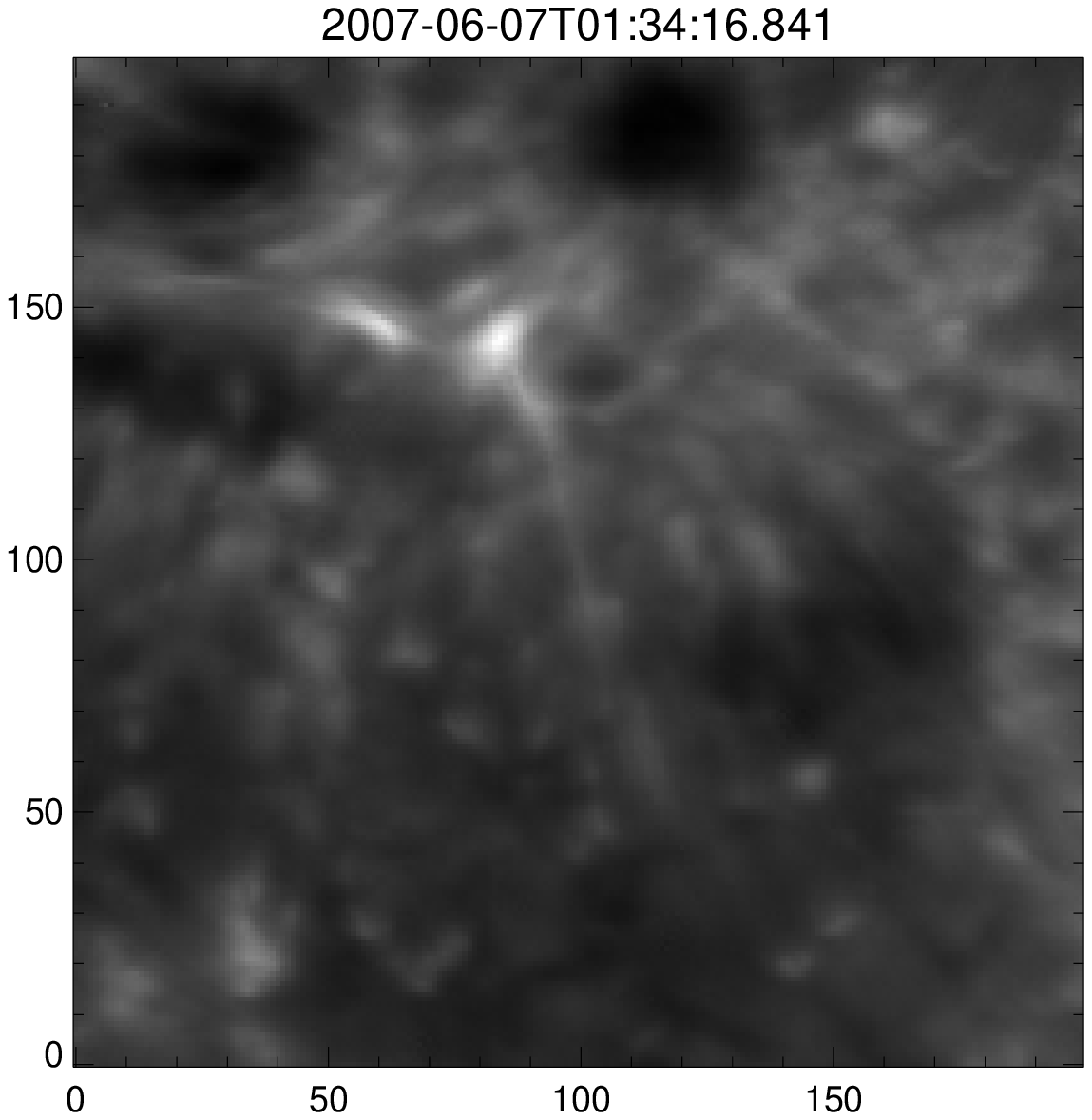} \\
\end{center}
\caption{The temporal evolution of the PLFs is shown in sequence of G-band and Ca~II~H images
alternatively. The date and time of the image acquisition is shown on the top of each image. The
numbers on the axes represent the pixel.}
\label{fig:3}
\end{figure*}

The evolution of the PLFs at the photospheric and chromospheric level is depicted in
Figure \ref{fig:3}. The PLFs in G-band and Ca~II~H images are shown alternatively for
near simultaneous time intervals.  On June 06, at 16:00~UT the PLFs is clearly visible in the
photospheric as well as in the chromospheric images. The PLFs appears to have very
small pore at one end and, however, compared to the size of PLFs it is very tiny, so the 
origin of PLFs cannot be this tiny pore. Later, at 00:30~UT on June 07, 2007 it reduced its 
size drastically and at 01:30~UT a small remnant of it is visible in the photospheric images. 
A similar type of evolution is observed in the chromospheric
images, except that some brightening can also be seen close to the PIL on June 7, 2007.

\begin{figure*}
\begin{center}
\includegraphics[width=100mm]{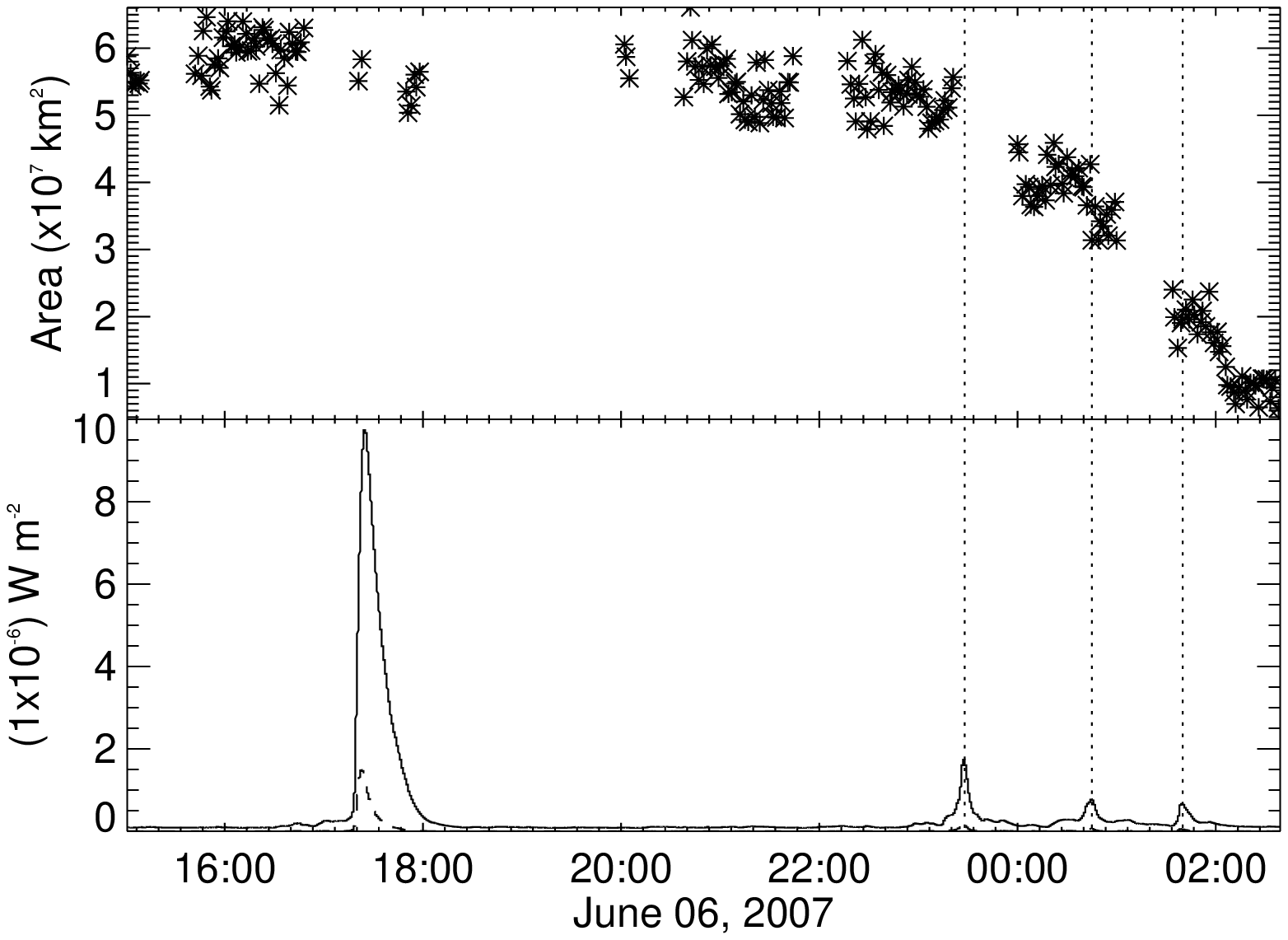} \\
\end{center}
\caption{Top: The area of the PLFs is plotted as a function of time.
Bottom: The GOES X-ray flus is plotted as a function of time.
The dotted vertical lines from left to right side represent the peak time of
C1.7, B7.6 and B6.6 class flares respectively.}
\label{fig:4}
\end{figure*}

\subsection{Temporal correlation}
We then measured the area of the PLFs quantitatively starting from 15:00 UT
of June 06, 2007. The area of the PLFs is measured from the G-band images.
We adopted the following methodology to measure the area of the PLFs. We
first smooth the image by 0.9$^{\prime\prime}$$\times$0.9$^{\prime\prime}$ 
pixel box to remove any small scale intensity variations in the time sequence of images. 
This small scale variations in the intensity of images could occur due to  
residual oscillations after filtering and 
residual flat-fielding errors. We then pick the pixels whose values smaller than
 0.6 times the value of the quiet sun intensity in the photosphere.
This criteria is same for all the images. This
method clearly picks the dark regions that contain not only the PLFs but also
the pore regions. To select only the PLFs region we then labeled the
regions which assigns a unique number to the individual detected sub-regions.
We use
`label$\_$region.pro' function of IDL data analysis platform to do the labeling.
Using this technique we isolated the PLFs  from rest of the region.  Once
the PLFs region has been isolated, by counting the number of pixels we estimated the 
area of the PLFs. The variation  of PLF's area
as a function of time is shown in Figure \ref{fig:4}. We have also plotted the 
GOES X-ray flux below the areal plot for easy comparison. In the plot, we
have drawn the vertical lines to show the peak time of the three flares
which are observed close to the PIL. We
notice the following features in the evolutionary trend:
(i) The PLFs area is almost constant, about 6$\times$10$^{7}$~km$^{2}$ before the
C1.7-class flare (first flare), that is, between 22:13 to 23:31 UT on June 06, 2007.
(ii) The PLFs area started to decrease during the C1.7-class flare and it
reduced to almost 3/4th of its original size during the B7.6 class flare (second flare).
(iii) After the B6.6-class flare (third flare) the size of the PLFs reduced drastically
and eventually the entire PLFs disappeared.
Thus, from these observations it is clear that the decrease of PLFs area is 
co-temporal with the duration of three recurrent flares which are closely 
spaced in time and located along the same PIL. The PLFs disappeared in about 
2~hours after it starts to decay.
There are  data gaps in between the observations, however a decreasing trend in 
PLFs area can be easily recognized. In section 3.4 we also study the spatial 
correlation between the flare brightening and the location of PLFs using 
co-aligned RHESSI X-ray images. These observations suggest  co-spatiality 
between these PLFs and the flare emission in X-ray. We discuss the
possible relation in section 4.

\subsection{Flow fields in the PLFs}
\begin{figure*}
\begin{center}
\includegraphics[width=70mm]{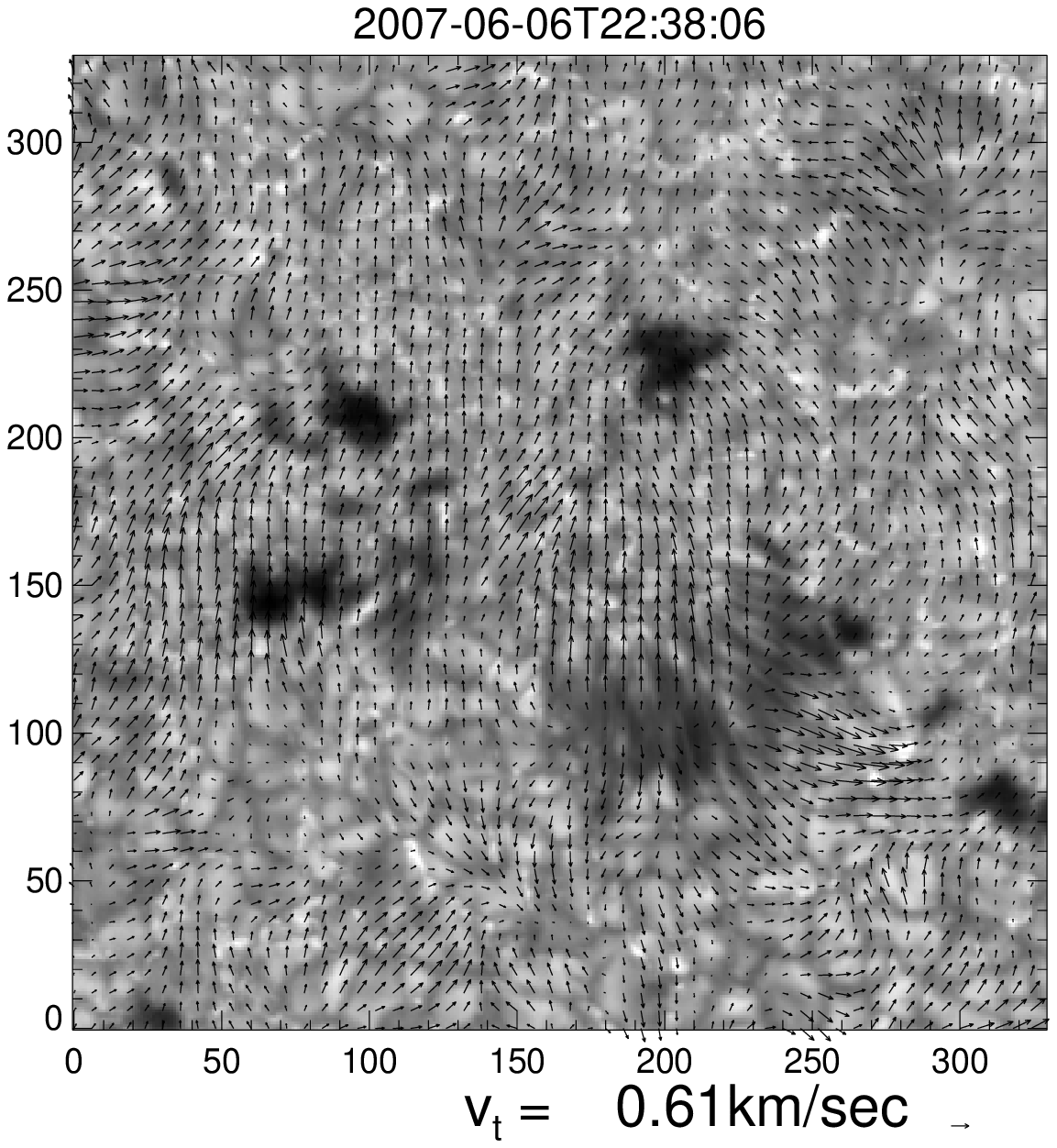}\includegraphics[width=70mm]{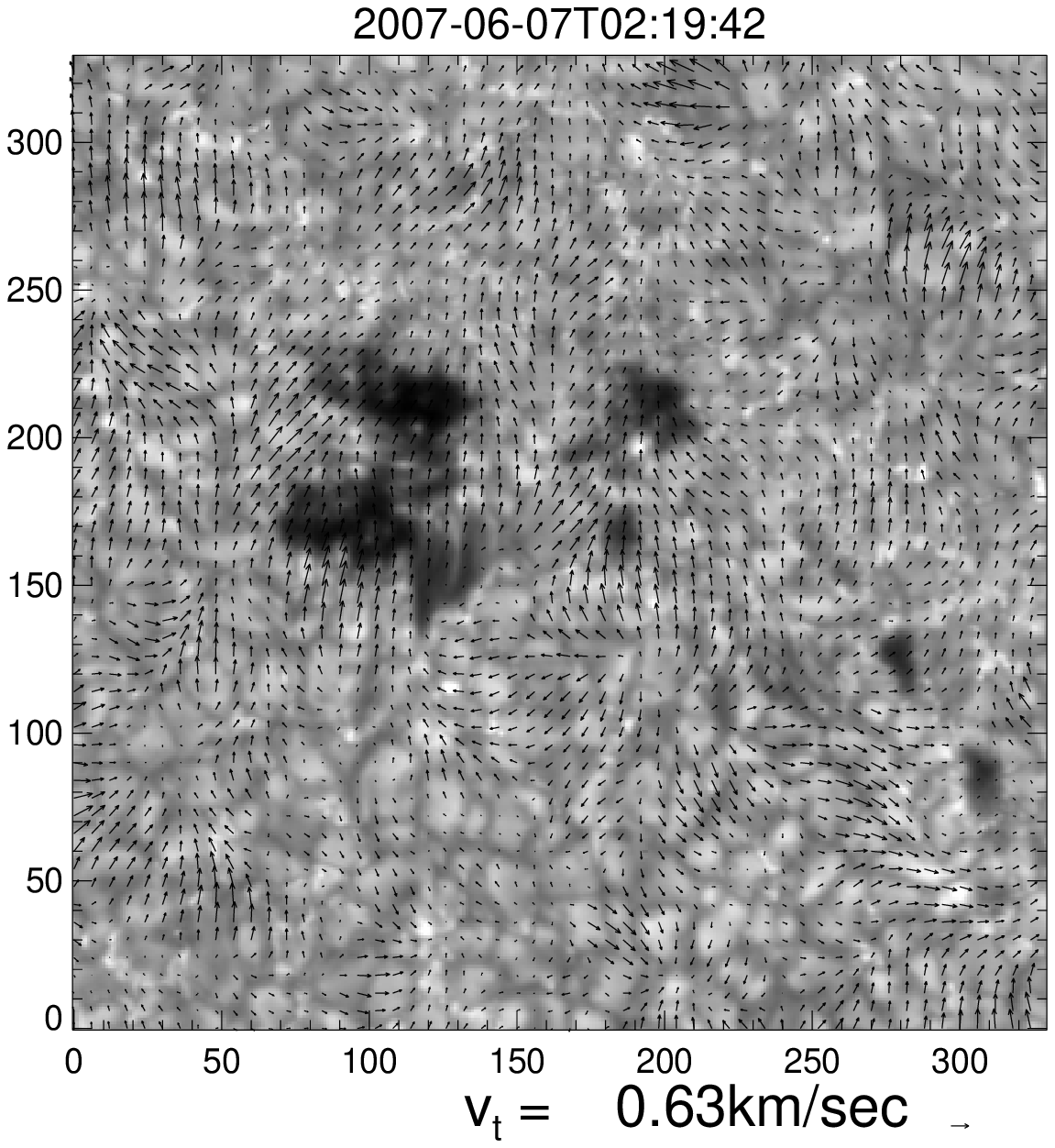} \\
\end{center}
\caption{The horizontal velocity vectors overlaid upon G-band image showing the PLFs.
Left: Image acquired before the C1.7 class flare. Right: Image acquired after the PLFs decay.}
\label{fig:5}
\end{figure*}

\begin{figure*}
\begin{center}
\includegraphics[width=70mm]{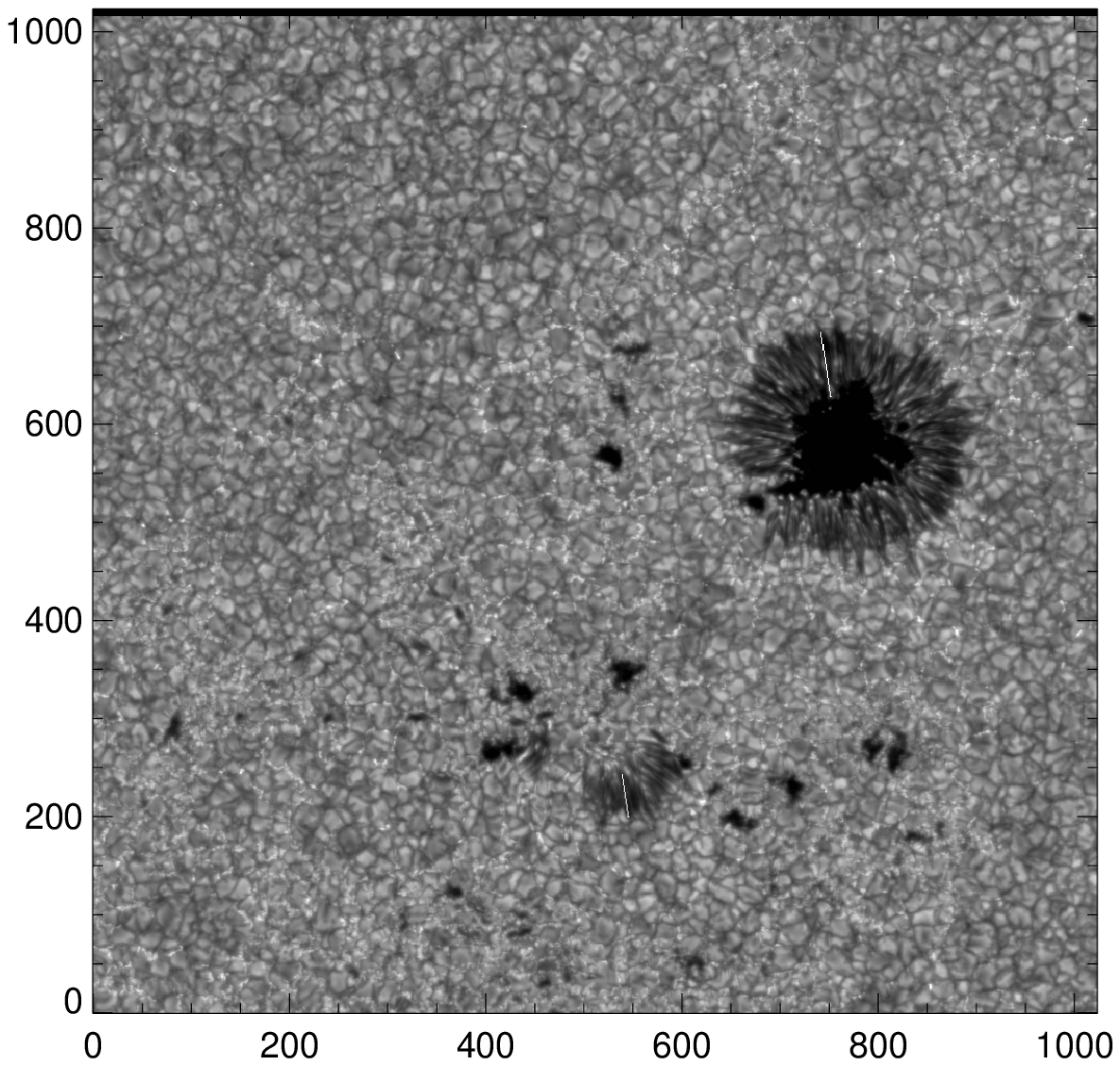}\includegraphics[width=70mm]{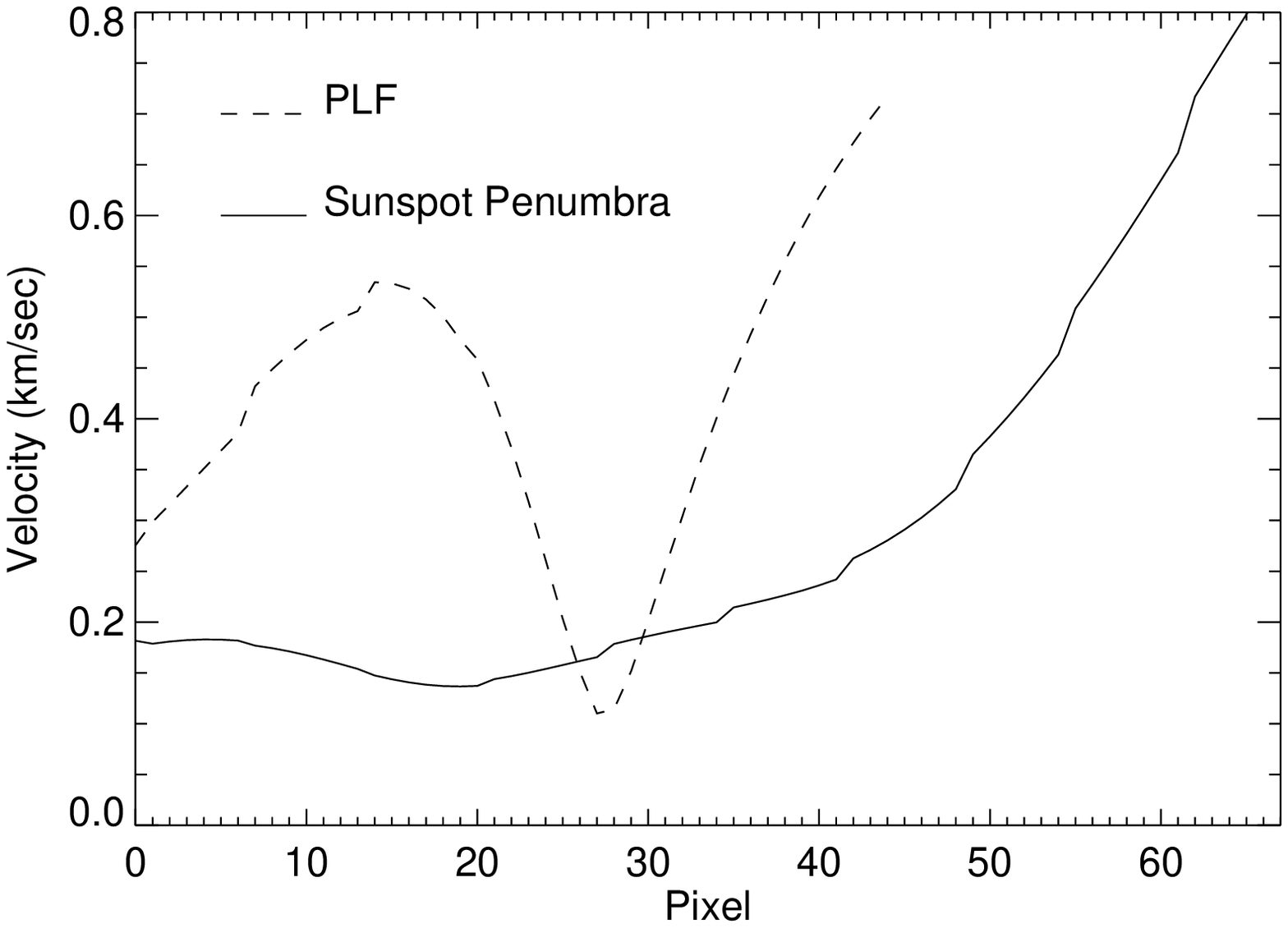} \\
\end{center}
\caption{Left: The location of the velocities extracted for the penumbra associated 
with the sunspot and the PLF is shown with white line. Right: The velocity for 
penumbral region and PLF region is plotted.}
\label{fig:6}
\end{figure*}

The penumbrae associated with the sunspots exhibit an outflow and inflow that appear 
to be originate from the mid portion of the penumbra (Ravindra, 2006). In general there 
is a radial outflow pattern around sunspots which is also known as the moat flow. 
From high resolution observations it was found that these radial outflows in the moat 
region appear only when penumbrae are present and are not seen
in those parts of irregular sunspots where penumbrae are missing 
(Vargas Domnguez et~al. 2007). In order to examine the flows in and around the 
PLFs we computed the horizontal flow fields using the
local correlation tracking (LCT: November and Simon, 1988) technique. 
Figure \ref{fig:5} shows the flow fields in the PLFs region. The left side image 
shows the flow field on June 06, before the
C1.7 class flare (first flare) and the right side image is obtained after the 
B6.6 class flare (third flare). In the left side image, the PLFs exhibits an outward 
flow which appears to be originating in the mid portion of the PLFs. The flow is in 
general, directed away from the PLFs, a little similar to what one finds around sunspots. 
However, here the PLFs is not in a radial symmetry
like sunspots and hence the associated flows are also non-symmetric. On the right side 
image, we notice that at the location where the PLFs disappeared, we can observe a 
supergranular like outward flow pattern which is roughly of the same size as that of the PLFs.

In order to compare the flow fields in the PLFs with the penumbral filaments 
we selected one region on the PLF and another on the sunspot penumbra, as shown in 
Figure \ref{fig:6}(left) with a white line. The LCT velocities are compared for both 
the regions and is shown in Figure \ref{fig:6}(right). The velocity in the penumbra 
which is directed towards the umbra is small, about 0.2~km~s$^{-1}$ and after the mid 
penumbra towards the outer penumbra the velocity increases. This result is similar to 
the observations of Sobotka, Brandt and Simon (1999).
The velocity is large near the outer penumbra, about 0.8~km~s$^{-1}$. On the other hand 
the velocity in PLFs increases on either side of mid portion of the PLF and it is minimum 
of about 0.2~km~s$^{-1}$ in the mid portion of the PLF. This suggests that there are 
diverging flows from the middle portion of PLFs. Clearly then there is a difference in the 
transverse field flow patterns in the PLF with respect to penumbra
associated with sunspot.

\subsection{Spatial correlation}
\begin{figure*}
\begin{center}
\includegraphics[width=70mm]{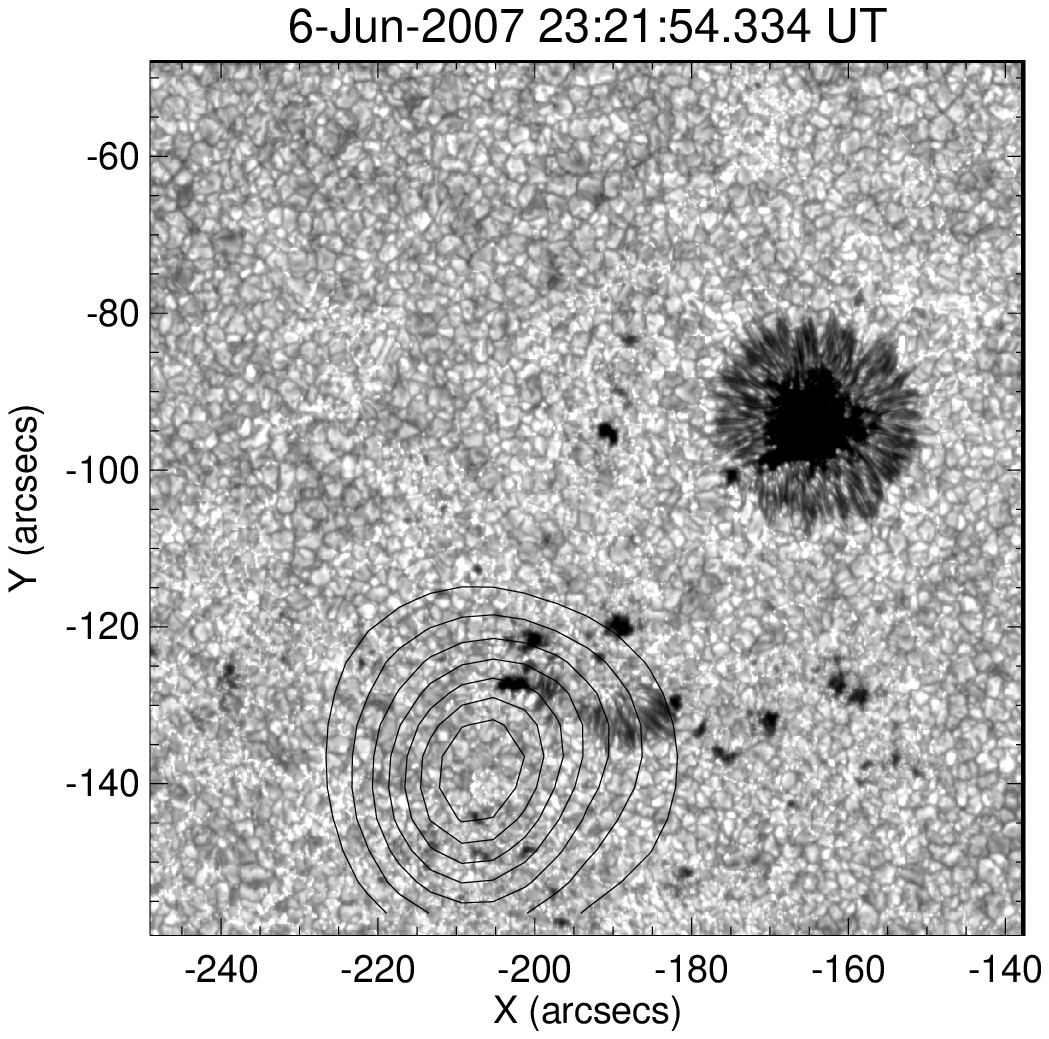}\includegraphics[width=70mm]{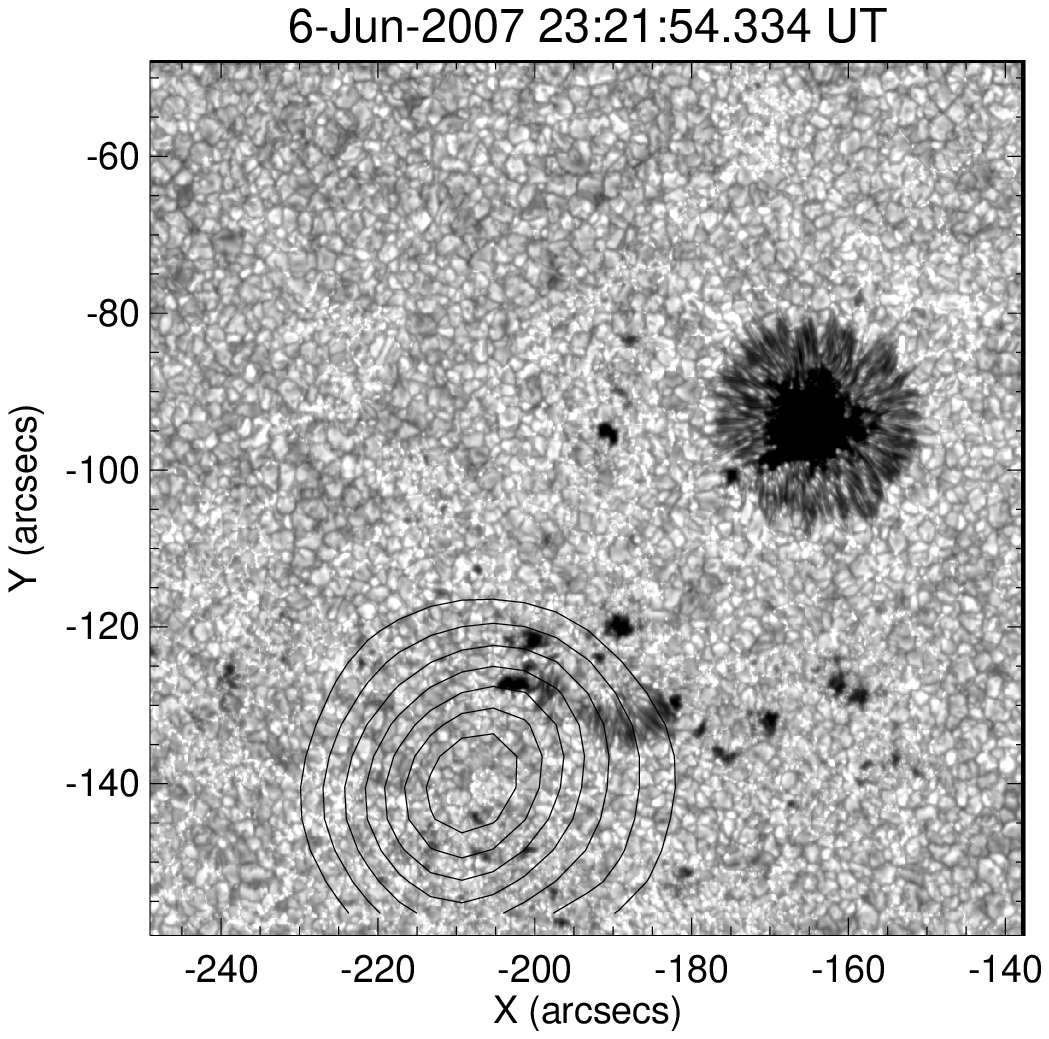} \\
\end{center}
\caption{The contours of the RHESSI hard
X-ray images overlaid upon the G-band images. Left: The contours of 6 to 12 keV and
Right: 12 to 25 keV.}
\label{fig:7}
\end{figure*}

Hard X-ray flux is the indicator of the footpoint of the source of
flare ribbons in the chromosphere (Masuda, Kosugi, and Hudson, 2001). It also
represents the thermal and non-thermal footpoints of the loops during 
micro-flares (Hannah, Krucker, Hudson, Christe, and Lin, 2008). We overlaid the 
contours of hard X-ray fluxes in the energy range from 6-12 and 12-25 keV from RHESSI 
on to the G-band images. As the pointing is known to be good for the
MDI, the G-band images were co-aligned with the MDI intensity images
by overlying the contours of MDI intensity image on the G-band image. The
contours of hard X-ray from RHESSI is overlaid on the co-aligned HINODE G-band
images (Figure \ref{fig:7}). This kind of low resolution RHESSI contours
overlaid upon the hi-resolution G-band and SOT/NFI images has been successfully 
used to locate the footpoints of the reconnecting loops in the corona 
(e. g. Matthews, Zharkov, and Zharkova, 2011; Watanabe et~al. 2010).
The left side image shows the contour maps for the
6-12 keV energy range and the right side is for 12-25 keV range.
These contours were plotted for the 40, 50, 60, 70, 80 and 90\% of the
maximum counts in the corresponding hard X-ray images. An observation
of these contour maps show that the contours are mainly concentrated near the
region of PLFs suggesting the spatial correlation between the flare location and the PLFs.

\subsection{Vector Field Maps}

\begin{figure*}
\begin{center}
\includegraphics[width=60mm]{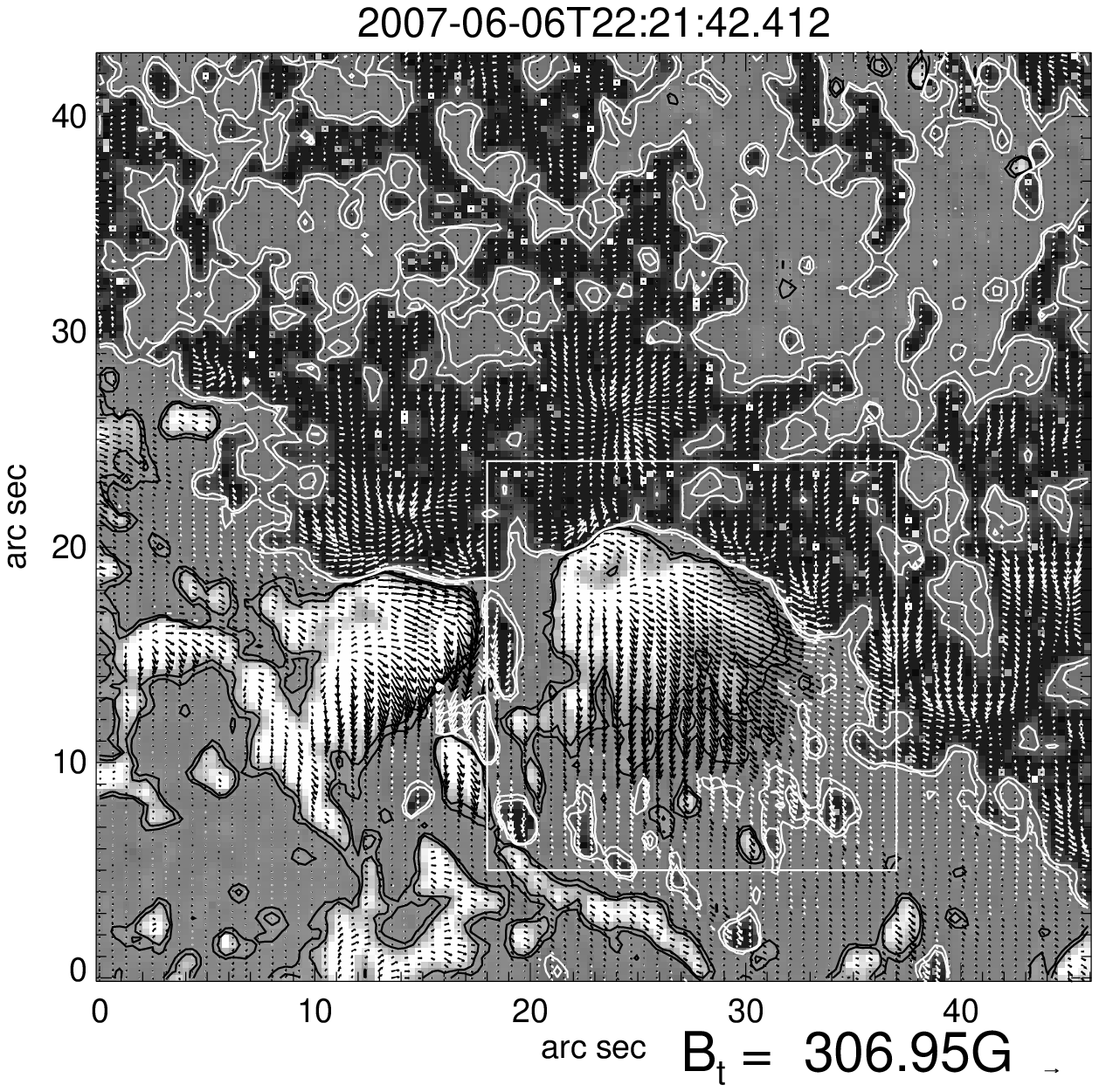}\includegraphics[width=60mm]{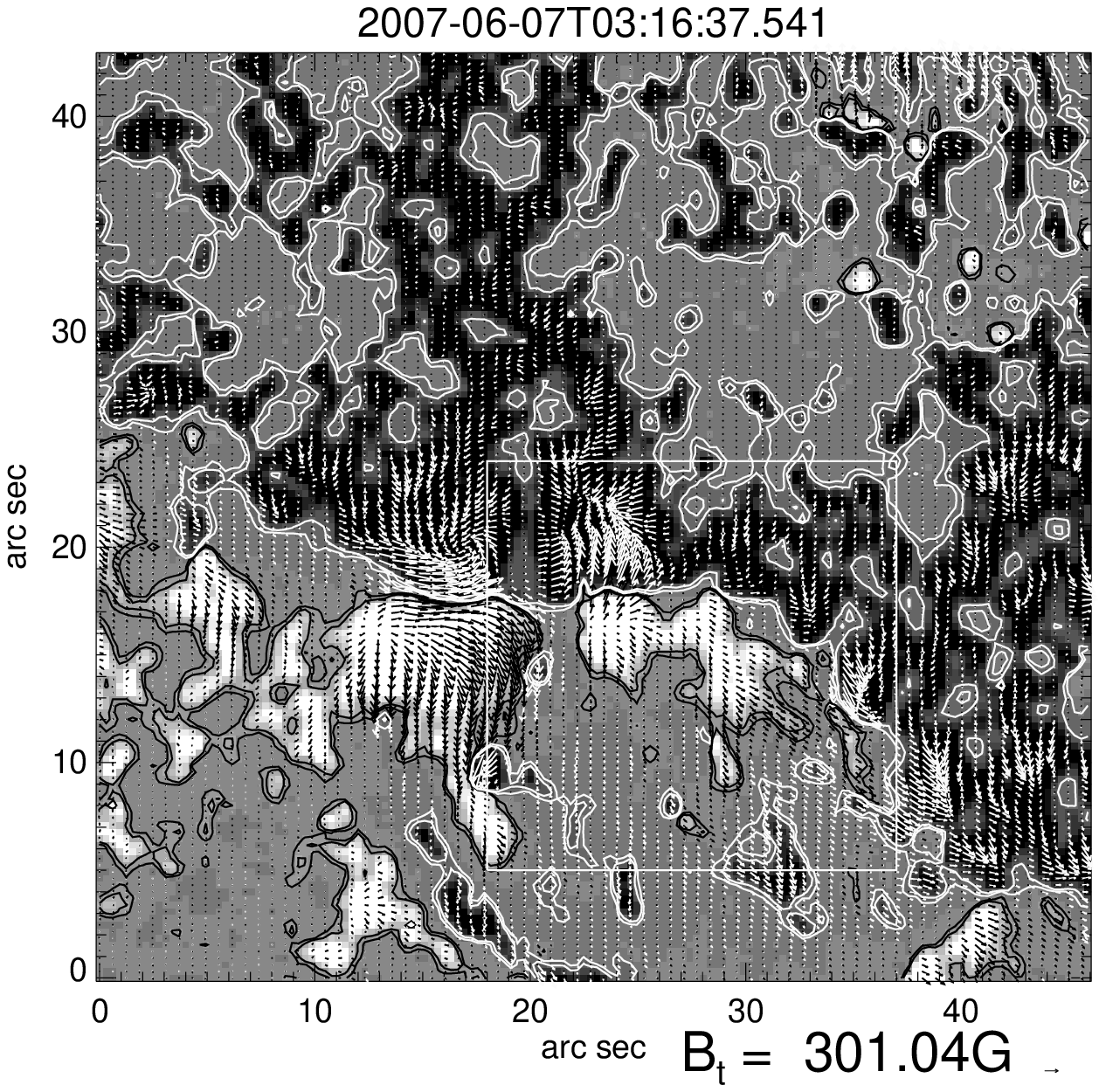} \\
\end{center}
\caption{ Transverse field vectors overlaid upon the vertical component of the magnetic
field. Left side images are obtained before the C1.7 class flare and the right side images 
are obtained after the PLFs decay.}
\label{fig:8}
\end{figure*}
Figure \ref{fig:8} (left) and (right) shows the vector magnetic fields before and 
after the flares, respectively.  The vector maps show an overlay of the horizontal 
component of the magnetic field (represented by arrows, where the length of the arrow 
is proportional to the magnitude of the horizontal field and the direction tells the 
direction of the horizontal field) upon the vertical magnetic fields (represented as 
gray scale image with overlaid contours).
In the map, the boxed region shows the PLFs location  and
the contours are drawn at $\pm$ 50 and 100~G levels of the B$_{z}$ component
of the magnetic fields. Comparison shows that the B$_{z}$ and B$_{t}$ component 
of the magnetic field in the PLFs location decreased substantially after the flare.

\begin{figure*}
\begin{center}
\includegraphics[width=70mm]{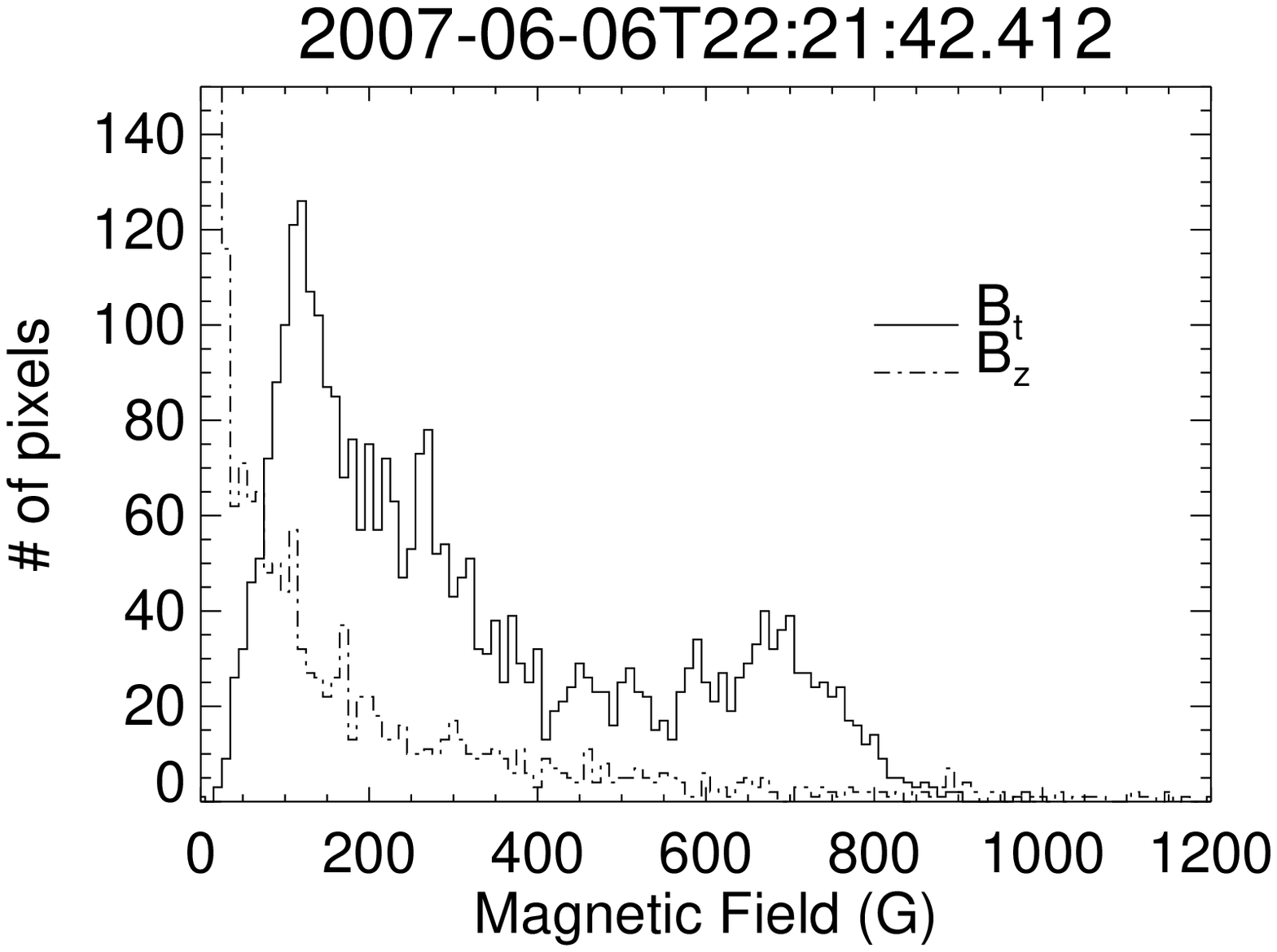}\includegraphics[width=70mm]{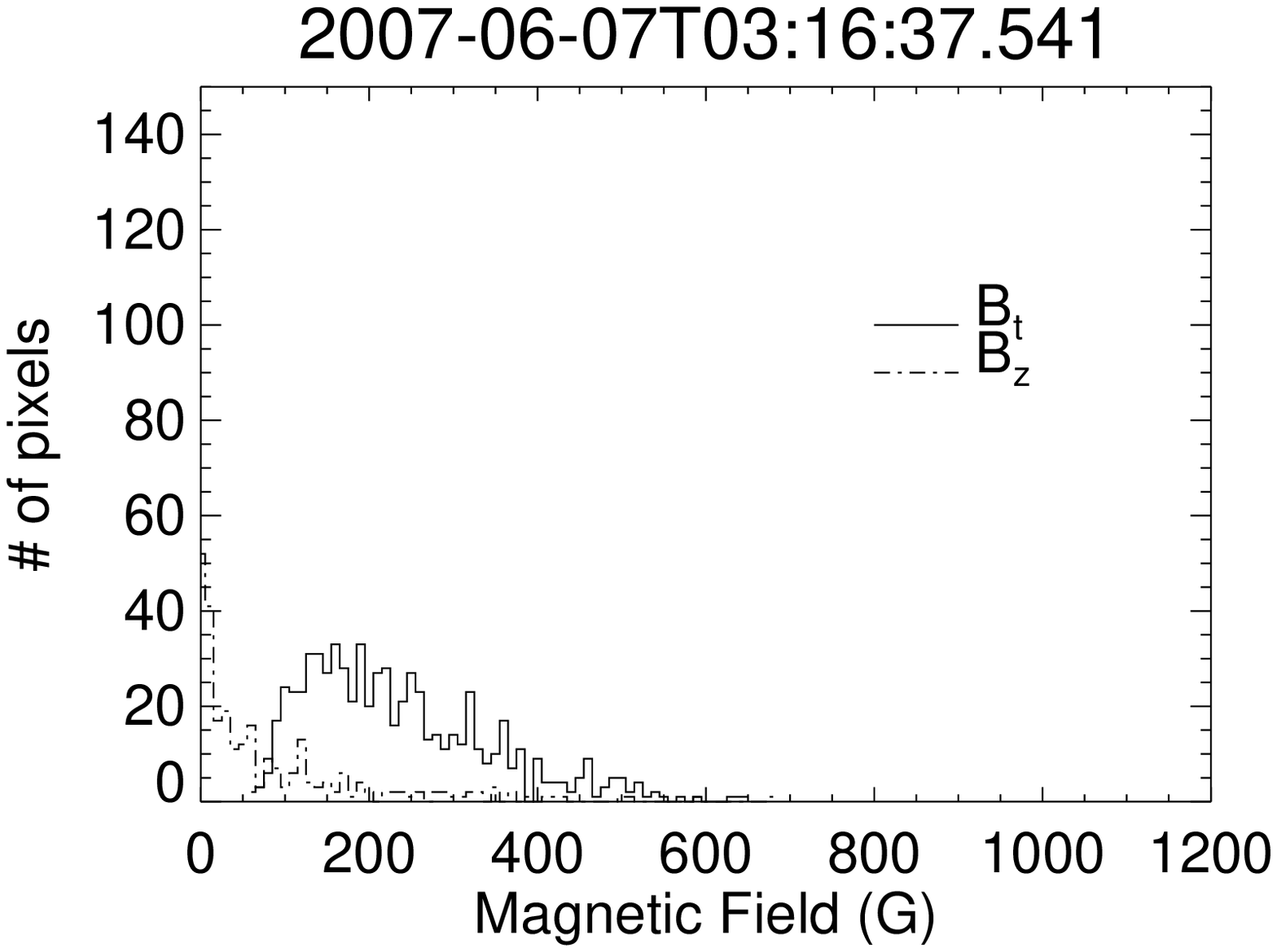} \\
\end{center}
\caption{Left: Histogram of the vertical component (B$_{z}$)
and transverse component of the magnetic fields (B$_{t}$) are plotted for the PLFs
obtained an hour before the flare. Right: Same as the left side plot, but for the
data obtained about  two hours after the flare. The thick line corresponds to
the histogram of transverse field strength and the dash dotted  line corresponds to the
histogram of vertical magnetic field strength.}
\label{fig:9}
\end{figure*}

Figure \ref{fig:9} shows the histogram representing the distribution of
B$_{z}$ and B$_{t}$ components of the magnetic fields in the region of PLFs.
The left and right side plots show the distributions before and after the
flare. The plot shows that a large amount of B$_{t}$ component is present
in PLFs before the flare and it decreased substantially after the flare. A similar
result is found for the B$_{z}$ component of the magnetic fields. Before
the flare the B$_{z}$ and B$_{t}$ components of the field in PLFs extended its tail
up to 1200~G. After the flare its strength has been reduced to less than half
of its value. This suggests that the magnetic field itself has decreased in this region.

\section{Discussion}
We have focused on a new type of  magnetic features in the photosphere near the 
PIL of a flaring active region, which we call as penumbra-like features (PLFs). These 
PLFs could be  observed due to high spatial resolution of Hinode SOT. These features are 
interesting in the sense that they are not associated with a pore or sunspot. It is 
generally seen that as pores grow larger they develop penumbrae abruptly, sometimes  
called rudimentary penumbrae which eventually develop as full sunspot penumbrae if the 
magnetic flux in the pore continues to grow. More detailed study of these features 
utilizing high-resolution imaging as well as spectro-polarimetery would shed more light 
on the formation and evolution of these features along with the magnetic and thermodynamic 
properties of these features. Also, it is important to know what is the vertical structure 
of these features, are these shallow or deep rooted, do they harbor similar Evershed flows 
and show similar uncombed magnetic structure as in sunspot penumbrae? Such a study is 
currently underway using Hinode SOT/SP and FG instruments and we plan to present the 
detailed observational characteristics of these features in a separate paper.

Here we focused on the disappearance of the PLFs in association
with the occurrence of three recurrent flares near the PIL where PLFs were located. 
The PLFs area was about 6$\times$10$^{7}$~km$^{2}$ initially and reduced rapidly during 
the interval of three flares suggesting a correlation between the occurrence of the flare
and the decay of the closely located PLFs.  The location of the hard X-ray flux obtained 
from the RHESSI coincides with the location of the PLFs suggesting that the footpoints of 
the reconnecting loops in the corona are rooted in or close to the PLFs.
The vector magnetic field associated with PLFs shows that these are largely transverse 
magnetic fields. A comparison of vector magnetic field retrieved by the Stokes inversion
of SP data before and  after the flares shows that there is
a substantial decrease in the transverse as well as vertical field component at the 
location of the PLFs.

Due to the lack of long time sequence of observations  it is difficult to tell exactly  
how the PLFs were formed. We can think of two possibilities: (i) Since PLFs are found in 
the location of  decaying sunspot it could be the remnant of the sunspot wherein the 
sunspot got disrupted and a cluster of penumbra got separated from the decaying sunspot. 
However, this picture does not conform to the typical observations of sunspot decay, 
where the sunspot looses it penumbral area gradually becoming so-called naked sunspot 
and resembling a pore, while the penumbrae are lost completely i.e., they do not form 
PLFs in general,  (ii) It could also happen that these are formed independently due to 
clustering of highly inclined fields near the PIL. Generally the PILs are observed to 
have strong horizontal fields and often harbor twisted fields (as evidenced in twisted 
filament structures lying along PIL. The twisted morphology becomes obvious when they 
erupt as helical structures). At present we can not confirm or rule out either of these 
possibilities, however, we hope future high resolution observations would help us to 
gain more knowledge about the formation of the PLFs.

The decay of PLF after the flare could be due to two main reasons, (i) change in field 
inclination from horizontal to vertical, and (ii) flux cancellation and/or submergence. 
In the former case we expect the vertical magnetic flux to increase after the flare, 
which is not observed in the vector magnetograms. Also, if PLFs field becomes vertical 
then we should see appearance of pores at the location of PLFs, which is not seen in 
the images. Thus, we can rule out this possibility.  The vector magnetic field 
observations by Hinode SOT/SP however, show a substantial decrease in the magnetic 
field in the PLFs area, which suggests that the decay of PLFs could be due to the 
flux cancellation and/or submergence. Since, we do not have a continuous vector magnetic 
field measurement it is very difficult to conclude firmly upon this. A future continuous 
high resolution data is therefore required.

In the past the disappearance of penumbra have been observed in large
X and M-class flares. The high-resolution
Hinode images gave us opportunity of observing the changes in the
photosphere even for the small flares with small penumbra like regions (PLFs). We 
could detect the changes in the area of the decaying PLFs during flare interval, but there 
is still a lack of vector magnetic field data at high temporal cadence to make more detailed 
study of the evolution of magnetic field in PLFs. Vector magnetograms from recently 
launched Solar Dynamic Observatory hold promise to the studies of vector magnetic field
evolution of such small features.

\section{Acknowledgments}
We thank the anonymous referee for constructive suggestions and comments on the paper.
Hinode is a Japanese mission developed
and launched by ISAS/JAXA, collaborating with NAOJ as a domestic partner, NASA and
STFC (UK) as international partners. Scientific operation of the Hinode mission is conducted
by the Hinode science team organized at ISAS/JAXA. This team mainly consists of scientists
from institutes in the partner countries. Support for the post-launch operation is provided
by JAXA and NAOJ (Japan), STFC (U.K.), NASA, ESA, and NSC (Norway).

\end{document}